\documentclass[pdftex,onecolumn,epjc3]{svjour3}
\pdfoutput=1
\usepackage{pstricks}
\usepackage{color}
\usepackage{soul}
\usepackage{placeins}
\usepackage{wrapfig}
\usepackage{tikz}
\usepackage{pst-node}
\usepackage{etex}
\usepackage{tikz-cd}
\usepackage{subfig}
\usepackage{graphicx}
\usepackage{bm}
\usepackage{xcolor}
\usepackage{amsmath}
\usepackage{amssymb}
\usepackage{mathtools}
\usepackage{mathrsfs}
\usepackage{amsfonts}
\usepackage{enumerate}
\usepackage{array}
\usepackage{booktabs}
\usepackage{textcomp}
\usepackage{caption}
\usepackage{cancel}
\usepackage{units}
\usepackage{multirow}
\usepackage{pst-all}
\usepackage{longtable}
\usepackage{relsize}
\usepackage{amsfonts}
\usepackage[T1]{fontenc}
\usepackage[utf8]{inputenc}
\numberwithin{equation}{section}

\usepackage{etoolbox}
\usepackage{enumitem}

\journalname{Eur. Phys. J. C}
\usepackage[margin=0.8in]{geometry}

\usepackage{calc}

\newlength{\depthofsumsign}
\newcommand{\bea}{\begin{eqnarray}\displaystyle}
\newcommand{\eea}{\end{eqnarray}}

\newcommand{\nn}{\nonumber \\}

\begin{document}
\setlength{\abovedisplayskip}{0pt}
\setlength{\belowdisplayskip}{0pt}
\title{Degeneration of Topological String partition functions and Mirror curves of the Calabi-Yau threefolds $X_{N,M}$}
%\author{M.Nouman Muteeb
%\thanksref{e1}
%}
%\thankstext{e1}{e-mail: nouman01uet@gmail.com}
%\keywords{Topological string partition function,$X_{N,M}$ mirror curve, degeneration,}
\author{Ambreen Ahmed \thanksref{e1}, M. Nouman Muteeb\thanksref{e2}
}
\thankstext{e1}{ambreen.ahmedgcu@yahoo.com}
\thankstext{e2}{nouman01uet@gmail.com}
\institute{Abdus Salam School of Mathematical Sciences, Lahore, Pakistan}

\maketitle
\begin{abstract}
In this article we study certain degenerations of the mirror curves associated with the Calabi-Yau threefolds $X_{N,M}$, and the effect of these degenerations on the refined topological string partition function of $X_{N,M}$. We show that when the mirror curve degenerates and become the union of the lower genus curves the corresponding partition function factorizes into pieces corresponding to the components of the degenerate mirror curve. Moreover we show that using degeneration of a generalised mirror curve it is possible to obtain the partition function corresponding to $X_{N,M-1}$ from $X_{N,M}$.
\end{abstract}
\keywords{refined topological string partition function, Calabi-Yau $X_{N,M}$ mirror curve, degeneration}
\tableofcontents
\newpage
%%%%%%%%%%%%%%%%%%%%%%%%%%%%%%%%%%%%%%
\section{Introduction: Refined topological strings on $X_{N,M}$ and corresponding Mirror Curves }\label{intro}
The non-compact Calabi-Yau  threefold (CY threefold) $X_{N,M}$ with $N, M \in \mathbb{N}$ \cite{Hohenegger:2016eqy,Bastian:2017ary,Ahmed:2017hfr,Haghighat:2018gqf,Hohenegger:2013ala,Hohenegger:2016yuv,Hohenegger:2015btj,Deger:2018kur}  has the structure of a double elliptic fibration with an underlying $SL(2,\mathbb{Z})\times SL(2,\mathbb{Z})$ symmetry. One elliptic fibration has the Kodaira singularity of type $I_{N-1}$ and the other elliptic fibration has $I_{M-1}$ singularity. The topological string partition function on $X_{N,M}$ was computed in \cite{Hohenegger:2016eqy} and shown to be related to the Little string theories (LSTs) with eight supercharges.  In the decompactification limit the low energy description of circle compactified  LSTs of  types $(M,N)$ and $(N,M)$ are described by quiver gauge theories with gauge groups $U(M)^N$  and $U(N)^M$ respectively.
In the geometric engineering argument  the M-theory compactification on a non-compact  Calabi-Yau threefold Y is described at low energies by the 5d $\mathcal{N}=1$ SCFTs. These SCFTs are  UV completions of the gauge theories we are interested in. The low energy gauge theory is completely specified by the requirement of  supersymmetry, once the gauge group $G$, hypermultiplet representation $R$ and the 5d Chern-Simons level $k$ is fixed. In taking the QFT limit the gravitational interactions are tuned off. This is achieved by sending the volume of $\mbox{Y}$ to infinity while keeping the volumes of compact four-cycles and two-cycles finite. This is equivalent to the  non-compactness condition of the CY threefold.  The coulomb branch of the  SCFT is identical to the extended K\"ahler cone of the threefold Y \cite{Bastian:2017ary, Jefferson:2018irk}.  The CY $Y$ can be understood as the singular limit of a smooth threefold $\widetilde{Y}$ in which certain number of  compact four-cycles have shrunk to a point. 
%The  existence of a gauge theory description of $\widetilde{Y}$ implies that the abelian gauge algebra is isomorphic to the quotient $H^2(\widetilde{Y},\mathbb{R})/H^2(\widetilde{Y},\mathbb{Z})$. The later enhances to a non-abelian gauge algebra in the singular limit $\widetilde{Y}\to Y$. 
The BPS states of the 5d theory correspond to M2-branes wrapping holomorphic two-cycles and M5-branes wrapping holomorphic four-cycles. The volume of the  two-cycles and four-cycles correspond to the masses of the BPS states. At a generic point of the Coulomb branch the two-cycles and four-cycles have non-zero volumes and the BPS spectra is massive. At the origin of the Coulomb branch some of the cycles may shrink to a point and indicate a local singularity  on the threefold.\\
The refined topological type IIA string partition function $\mathcal{Z}_{N,M}$ of $X_{N,M}$ can  efficiently be computed using the refined topological vertex formalism\cite{Iqbal_2009}. The partition function $\mathcal{Z}_{N,M}$ takes the form of an infinite series expansion. The expansion parameters depend on the choice of a preferred direction common to all vertices of the toric web diagram. Different choices of the preferred direction give equivalent but seemingly different  representations of   $\mathcal{Z}_{N,M}$  \cite{Bastian:2017ary,Hohenegger:2013ala,Hohenegger:2016yuv}. 
Lately another powerful method of computing the partition function was proposed in  \cite{Haghighat:2013gba} in terms of  M-strings, which are one dimensional intersections of M5 and M2 branes.
The  table given in figure \ref{figure:coordinates} summarises the coordinate labels  and specifies the world volume directions of BPS M5-M2-M-string configuration.
\begin{figure}
\begin{center}
\begin{tabular}{ |p{2cm}|p{7.3cm}|  }
 \hline
 \multicolumn{2}{|c|}{11d M-theory space-time} \\
 \hline
 & $x^0\quad x^1 \quad x^2\quad x^3\quad x^4\quad x^5\quad x^6\quad x^7\quad x^8\quad x^9\quad x^{10}$\\
 \hline
 M5-branes   & $\times\quad \times \quad \times\quad \times\quad \times\quad \times\quad \quad \quad \quad \quad $ \\
 M2-branes&   $\times\quad \times \quad \quad \quad \quad \quad \quad \quad \qquad\times \quad $ \\
 M-string &$\times\quad \times \quad \quad \quad \quad \quad \quad \quad \quad \quad $ \\
 \hline
\end{tabular}
\caption{coordinates of the 11d M-theory space-time}
\label{figure:coordinates}
\end{center}
\end{figure}
%\newpage
The M5-branes are separated along the  compactified $x^6\sim x^6+2\pi R_6$ dimension with the positions parameterised by scalars VEVs $\{a_1,...,a_{M}\}$ where $M$ denotes the total number of M5-branes and $a_i-a_{i+1}$ are the VEVs of the scalars of 6d tensor multiplets. The M2-branes are stretched between these M5-branes. For the transverse space $\mathbb{R}^4$ we can have only one stack of M2-branes between M5-branes. However it is possible to perform an orbifolding \cite{Haghighat_2014} of  the transverse  $\mathbb{R}^4$ such that the mass deformation and supersymmetry remain preserved. The orbifolding  allows the multiple stacks of M2-branes with each stack charged under the orbifold action. For the M-string dual to $(N,M)$ web diagram there will be $N$ stacks of M2-branes, with i-$th$ stack consisting of $k_i$ number of them. In gauge theory $k_i$  characterises the instanton number.
 It was shown subsequently in \cite{Hohenegger:2013ala} that the M-string partition function $\mathcal{Z}(N,M)$  is the generating function of the equivariant $(2,0)$ elliptic genus of the M-string world sheet,
\bea
\mathcal{Z}(N,M)=\sum_{\vec{k}}Q_1^{k_1}Q_2^{k_2}...Q_M^{k_M} \chi_{ell}(M(N,\vec{k}),V_{\vec{k}})
\eea
 Its target space is the product of moduli spaces of  $U(N)$ instantons of charge $k_i$ on $\mathbb{C}^2$ : $M(N,\vec{k}):=M(N,k_1)\times M(N,k_2)\times ...\times M(N,k_N)$ along with a vector bundle $V(N,M)$ on it. The mass deformation is taken care of by an extra $U(1)_m$ action with equivariant parameter $m$. The vector bundle is special in the sense that only right moving fermions couple to it. The moduli space $M(N,\vec{k})$ is  nothing other than the moduli space of M-strings.
\begin{figure}[hbt!]
        \center{\includegraphics[width=6in]{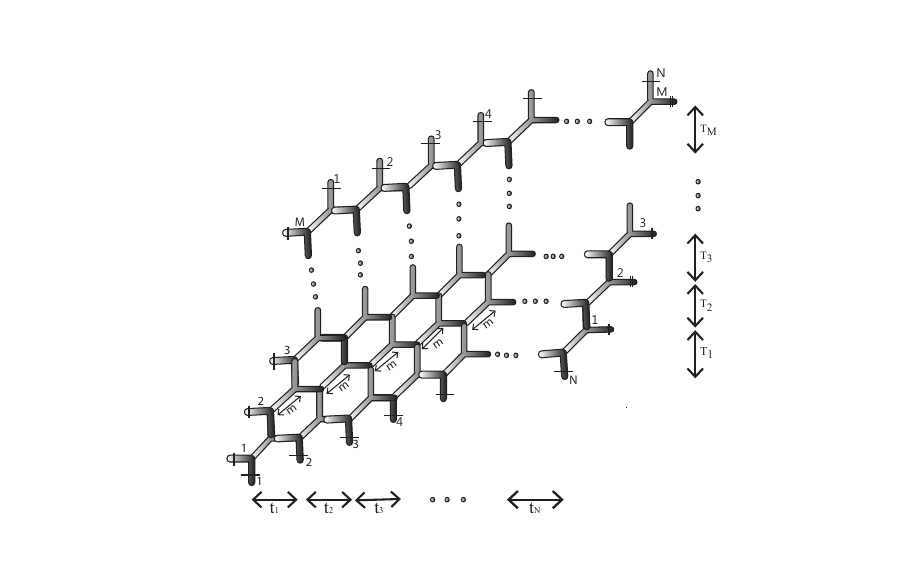}}
    %    \caption{\label{fig:X12} tessellation of Newton polygons and web diagram of $X_{1,2}$}
   \caption{Web diagram of $X_{N,M}$. $t_i\in \{t_1,...,t_N\}$ denotes the distance between $i$-th and $i+1$-th lines and $T_i\in \{T_1,...,T_M\}$ denotes the distance between $i$-th and $i+1$-th lines. $m$ denotes the K\"ahler parameter of the diagonal $\mathbb{P}^1$s. The double and single bars $|$ and$-$ indicate the periodic identifications.}
\label{Fig:WebToric3}
      \end{figure}

 For example the specific values $M=1,N=k$ correspond to a single M5-brane wrapped on parallel $S^1$ and $k$ stack of $M2$-branes wrapped on the transverse $S^1$ and ending on the M5-branes. The stack of M2-branes appear as coloured points in the $\mathbb{R}^4_{||}$ that resides inside the M5-brane world volume and transverse to the M-string world sheet. Thus for the configuration that involves $n_l$ number of M2-branes in the $l$-th stack, where $l=1,...,k$,  the moduli space is obviously the product of Hilbert scheme of points as follows
\bea
\mbox{H}:=\mbox{Hilb}^{n_1}[\mathbb{C}^2]\times \mbox{Hilb}^{n_2}[\mathbb{C}^2]\times\cdots\times \mbox{Hilb}^{n_k}[\mathbb{C}^2]
\eea
The vector bundle $\mbox{V}$ over $\mbox{H}$ that is required for $(2,0)$ world sheet theory has been determined in
\cite{Haghighat:2013gba} and turns out to be the following
\bea
V_I=\oplus_{t,s=1}^N \mbox{Ext}^1(\mbox{I}_r,\mbox{I}_s)\otimes L^{-\frac{1}{2}}
\eea
where $\mbox{I}=(\mbox{I}_1,\mbox{I}_2,...,\mbox{I}_N)\in H$. Roughly speaking Ext groups count the massless open string states for strings that are stretched between D-branes wrapped on complex submanifolds of CY spaces.    Note that each factor $ \mbox{Ext}^1(\mbox{I}_r,\mbox{I}_s)\otimes L^{-\frac{1}{2}}$ in the fibre denotes the contribution of a pair of stack of M2-branes ending on a single M5-brane from opposite sides.
 In other words there is an isomorphism between the degrees of freedom on the $(N,M)$ 5-branes web and the moduli space of M-strings, $\mbox{M}(N,\vec{k})$. Using equivariant fixed point  theorems one only needs to know the fibres of the bundle $\mbox{V}(N,M)$ over the fixed points.\\
 The  weights of $\mbox{V}(N,M)$ at the fixed points $\vec{I}^{(1)},\vec{I}^{(2)},...,\vec{I}^{(M)}$ are given by the following Chern character expansion \cite{Hohenegger:2013ala}
 \bea\label{eq:fixedpoint}
 \sum_{weights}e^w=\sum_{p=1}^M\sum_{r,s=1}^NQ_me^{i(a_r-a_s)}\bigg(\sum_{(i,j)\in\nu_r^{(p)}}t^{\nu_{s,j}^{t,(p+1)}-i+\frac{1}{2}}q^{\nu_{r,i}^{(p)}-j+\frac{1}{2}}+\sum_{(i,j)\in\nu_s^{(p+1)}}t^{-\nu_{r,j}^{t,(p)}+i-\frac{1}{2}}q^{-\nu_{s,i}^{(p+1)}+j-\frac{1}{2}}\bigg)\nonumber\\
 \eea
 where $\nu_1^{(1)},\nu_2^{(1)},...,\nu_N^{(1)};\nu_1^{(1)},...,\nu_N^{(1)}$ label the fixed points. The elliptic genus is then given as follows
 \bea
 Z
 %int_M\prod_i\frac{x_i\vartheta(\tau,\widetilde{x_i}+z)}{\vartheta(\tau,x_i)}
 =\int_M\prod_i\frac{x_i\theta_1(\tau,\widetilde{x_i}+z)}{\theta_1(\tau,x_i)}
 \eea
 where $x_i$ and $\widetilde{x}_i$  denote the Chern roots respectively of the tangent bundle and vector bundle $\mbox{V}(N,M)$ as can be read from  (\ref{eq:fixedpoint}) and the theta function of first kind $\theta_1(\tau,z)$ is  defined by
 \bea
 \theta_1(\tau;z)=-ie^{\frac{i\pi}{4}}(e^{i\pi z}-e^{-i\pi z})\prod_{k=1}^{\infty}(1-e^{2\pi i k\tau})(1-e^{2\pi i k\tau}e^{2\pi i kz})(1-e^{2\pi i k\tau}e^{-2\pi i kz}).
 \eea
  \\
 More succinctly, the Nekrasov partition function of the gauge theory on the D5-branes  of the web is identical to the appropriately normalised  topological string partition function of CY threefold $X_{N,M}$ and it is also the generating function of the $(2,0)$ elliptic genus of the product of instanton moduli spaces $\mbox{M}(N,\vec{k})$ on which the bundle $\mbox{V}(N,M)$ coupled to the right moving fermions exists.\\
 
 \subsection*{Presentation of the paper}
%In section 2 we summarise and elucidate some properties of the generating function of Gromov-Witten invariants of $X_{N,M}$ and its  modular properties. It is indicated that the Gopakumar-Vafa infinite product representation for
% $\mathcal{Z}_{X_{N,M}}$ can be obtained in some denigrate limits of the K\"ahler parameters.
  We summarised the type IIA/type IIB mirror symmetry conjecture in the introduction ($\ref{intro}$).
 In section (\ref{degenmirror}) we construct the quantum mirror curve of $X_{N,M}$ and study the limits in which it can be reduced to a lower genus curve. In  section  (\ref{splitdegn}) we show that in the splitting degeneration limit the partition function $\mathcal{Z}_{X_{N,M}}$ is recursively related to the partition function $\mathcal{Z}_{X_{N,M-1}}$ and we show this degeneration  pictorially. %In the last section (\ref{physicalcon}) we briefly mention some physical consequences of the degenerations discussed in the previous sections.
  In the appendix we reproduce the proof of an identity used in the main text.

\section{(p,q) webs and the mirror curves}\label{PQwebmirror}
%The CY threefold $X_{N,M}$ is a double elliptic fibration of type $A_{N-1}\times A_{M-1}$ over a non-compact base $\mathbb{C}$. It is toric with a web diagram shown in figure (\ref{Fig:WebToric3}), which is drawn on a torus with radii of the two circles  being dual to the K\"ahler classes of the elliptic fibers of $X_{N,M}$.  These threefolds were studied by  \cite{Kanazawa:2016tnt,Hohenegger:2016yuv,Hohenegger:2013ala,Hohenegger:2016eqy,Hohenegger:2015btj} as examples of toric varieties of infinite type.
%The toric CY threefold $X_{N,M}$ can be obtained by $\mathbb{Z}_N\times\mathbb{Z}_M$ orbifolds of $X_{1,1}$. This set up is dualizable to $(p,q)$ $5$-brane webs and realise various five and six-dimensional gauge theories. The 5-brane web is identical to the toric web underlying $X_{N,M}$.\\
 We can consider \cite{Aganagic:2001nx,Aganagic:2000gs} the A-model topological strings on a toric CY threefold $M=\mathbb{C}^{l+3}//U(1)^l$. Algebraically $M$ is defined by the following set of constraints
\bea\label{eq:c1}
\sum_{i=1}^{l+3}Q_i^{a}|X_i|^2=k^{a},\quad a=1,...,l
\eea
modulo the action of $U(1)^l$, where each $X_i$ parameterizes a complex plane $\mathbb{C}$ and can be visualised as $S^1$-fibrations over $\mathbb{R}_+$. In this way $M$, as defined by (\ref{eq:c1}), is a $T^3$-fibration over a non-compact convex and linearly bounded subspace in $\mathbb{R}^3$, with $T^3$ parametrised by $\{\theta_i\}$ coordinates. $k^a\in \mathbb{R}_+$ are called the K\"ahler parameters. The CY condition
\bea
c_1(TM)=0
\eea
holds iff
\bea\label{eq:c2}
\sum_{i=1}^{l+3} Q_i^a=0,\quad a=1,...,l
\eea
Inspecting equation (\ref{eq:c1}) makes it clear that since $Q_i^a\in \mathbb{Z}$, all toric CY threefolds are constrained to be non-compact. The second constraint (\ref{eq:c2}) furnishes a representation of $M$ as  $\mathbb{R}_+\times T^2$ fibered over $\mathbb{R}^3$. In this way the toric threefold M allows its construction by gluing patches of $\mathbb{C}^3$. \\
%The toric diaram $\Gamma_M$ corresponding to $M$ specifies the loci along which the $S^1$ fibers degenerate. The boundary of the region, denoted by $B$, is defined by $X_i=0$. For each value of $i$  this zero locus defines a 2-plane in $\mathbb{R}^3$ whose normal vector satisfies
%\bea
%\sum_{i=1}^{3+l}Q_a\vec{n_i}=0
%\eea
%Obviously, the $S^1$ parametrised by $\theta_i$ shrinks at $|X_i|=0$ and at the intersection of two such loci
%$S_{ij}=\{|X_i|=0 \}\cap \{|X_j|=0 \}$, two circles $\mathbb{S}^1$s shrink to zero size. For $S_{ij}$  a closed line in $\partial B$ the open $S^1$ bundle over it is a $\mathbb{P}^1$. For $S_{ij}$ a half open line it represents a non-compact direction $\mathbb{C}$. It is clear now that the relative position of $X_i, X_j$ is determined by the length of the line segment $S_{ij}$ which is nothing other than the K\"ahler parameter of the corresponding $\mathbb{P}^1$.
%The CY condition (\ref{eq:c2})  and the $\mathbb{T}^2$ fibration structure allows to project the $S_{ij}$s onto $\mathbb{R}^2$ in such a way that all the information about the geometry of $M$ is contained in it. Projecting all the $S_{ij}$s onto $\mathbb{R}^2$ in this way constitute the toric diagram $\Gamma_M$.\\
To construct the mirror N of the threefold M, consider variable $v_1,v_2 \in \mathbb{C}$, and the homogeneous coordinates $x_i=: e^{y_i}\in \mathbb{C}^*, i=1,...,l+3$ related to $X_i$ by $|x_i|=e^{-|X_i|^2}$. The variables
$x_i$ are constrained by $x_i\sim \lambda x_i$ for $\lambda\in \mathbb{C}^*$. The mirror geometry $N$ is then given by the algebraic equation
\bea\label{eq:mcy}
v_1 v_2&=&\sum_{i=1}^{l+3}x_i,\nonumber\\
\eea
subject to the constraints
\bea
\prod_{i=1}^{l+3}x_i^{Q_i^a}&=& e^{-r^a-i \theta_a},\quad a=1,...,l
\eea
All of these equations can be combined into a single equation
\bea
v_1v_2=h(x,y;r^a,\theta_a)
\eea
where $x,y\in \mathbb{C}^*$.  The function $h(x,y;r^a,\theta_a)$ can be decomposed into pant diagrams described by
\bea
e^x+e^y+1=0.
\eea
 The last equation describes a conic bundle over
$ \mathbb{C}^*\times \mathbb{C}^*$ in which the fibers degenerate over two lines over the family of Riemann surfaces $\Sigma:{g(x,y;r^a,\theta_a)=0}\in \mathbb{C}^*\times \mathbb{C}^*$.
If the toric diagram of $M$ is thickened, what emerges is nothing else but $\Sigma$ ; the genus of $\Sigma$ equals the number of closed meshes and the number of punctures equals the number of semi infinite lines in the toric diagram\footnote{It is a standard in literature to call $\Sigma$ the mirror curve.}.
%\subsection{Kodaira Spencer theory}
In the topological A-model the topological vertex computation can be interpreted as the states of a chiral boson on a three-punctured  sphere. This chiral boson on each patch of the sphere is identified with the Kodaira Spencer field on the Riemann surface embedded in the CY threefold of  mirror  topological B-model \cite{Huang:2011qx,Hellerman_2012,Gopakumar:1998jq,Gopakumar:1998ii,huang2010direct,katz1997mirror,Bhardwaj_2016}.
The A-model closed topological strings on toric CY threefold, with or without D-branes, is computable by gluing  cubic topological vertex expressions. On the mirror  B-model the gluing rules are equivalent to the operator formation of the Kodaira Spencer theory on the Riemann surface.The elliptic Calabi-Yau threefold  $ X_{N,M}$ is dual to the brane web of type IIB $M$ NS5-branes and $N$ D5-branes wrapped on two $\mbox{S}^1$s.
We denote by $\{y^0,y^1,y^2,y^3,...,y^9\}$ the coordinates of type IIB string theory vacuum $\mathbb{R}^{1,9}$. The common worldvolume of the 5-branes along $\{y^0,y^1,y^2,y^3,y^4\}$ gives rise to the gauge theory under consideration and the $(p,q)$ brane web is arranged in the $\{y^5,y^6\}$ plane which  is compactified to a torus $\mbox{T}^2$. The $(p,q)$-charges and their conservation encode the details of the five-dimensional mass deformed supersymmetric  gauge theory.
\\
The curve associated to a grid diagram is written as the zero locus of a  sum  of  monomials, with each monomial  associated to a vertex of the grid diagram. For example $A_{kl}X^kY^l$ is a monomial that corresponds to the vertex $(k,l)$. The modulus of the curve $A_{kl}$ is determined by imposing a set of condition: each link on the grid joining e.g. $(k,l)$ to $(u,v)$ uniquely corresponds to a link on the web, which is orthogonal to the former. If the link on the web is given by the line $py=qx+\alpha$, the orthogonality condition is expressed as
\bea
(k,l)-(u,v)=(-q,p)
\eea
and the constraint is given by
\bea
py=qx+\alpha:\quad A_{kl}=e^{\beta\alpha}A_{uv}
\eea
\\
In other words the mirror curves of toric CY threefolds are determined by the corresponding Newton polygons. The line in the web \cite{Aharony_2000,nekrasov2002seibergwitten,Iqbal_2009,Aganagic_2004,Bershadsky_1996,leung1997branes,Aharony_1998} orthogonal to the line in the Newton polygon joining the coordinates,let's call them $(k_{1},\ell_{1})$ and $(k_{2},\ell_{2})$ and passing through the point $(x_{0},y_{0})$ is given by ,
\bea
(\Delta \ell) \,y+(\Delta k)\, x&=&(\Delta \ell)\, y_{0}+(\Delta k)\, x_{0}
\eea
where $\Delta \ell=\ell_{2}-\ell_{1}$ and $\Delta k=k_{2}-k_{1}$. Since the choice of $(x_0,y_0)$ is arbitrary, we get
\bea
(\Delta \ell)\, y+(\Delta k)\, x=\alpha
\eea
The equation of the Riemann surface in this patch is given by exponentiating and complexifying $(x, y)$ to $(u, v)$,
\bea
X^{\Delta k}\,Y^{\Delta \ell}=-e^{\widetilde{\alpha}} \, ,
\eea
where $X=e^{u}$ and $Y=e^{v}$ with $u,v\in \mathbb{C}$ and $\text{Re}(\widetilde{\alpha})=\alpha$. Since the imaginary part $\widetilde{\alpha}$ is not determined, we have introduced a factor of $-1$
% (shifting the imaginary part by $i\pi$)
 for later convenience. With this choice,  $\widetilde{\alpha}$ will be identified with the complexified K\"ahler parameters.
In the mirror curve,  we will have
\bea
A_{k_{1}\ell_{1}}X^{k_{1}}Y^{\ell_{1}}+A_{k_{2}\ell_{2}}X^{k_{2}}Y^{\ell_{2}}&=&0
\eea
which can be solved to give
\bea
X^{\Delta k}Y^{\Delta \ell}=-\frac{A_{k_{1}\ell_{1}}}{A_{k_{2}}\ell_{2}}~~~~~~~ \implies ~~~~~A_{k_{2}\ell_{2}}=A_{k_{1}\ell_{1}}\,e^{-\widetilde{\alpha}}
\eea
\section{Mirror curves and their degenerations}\label{degenmirror}
\begin{figure}[!htb]
        \center{\includegraphics[width=6in]{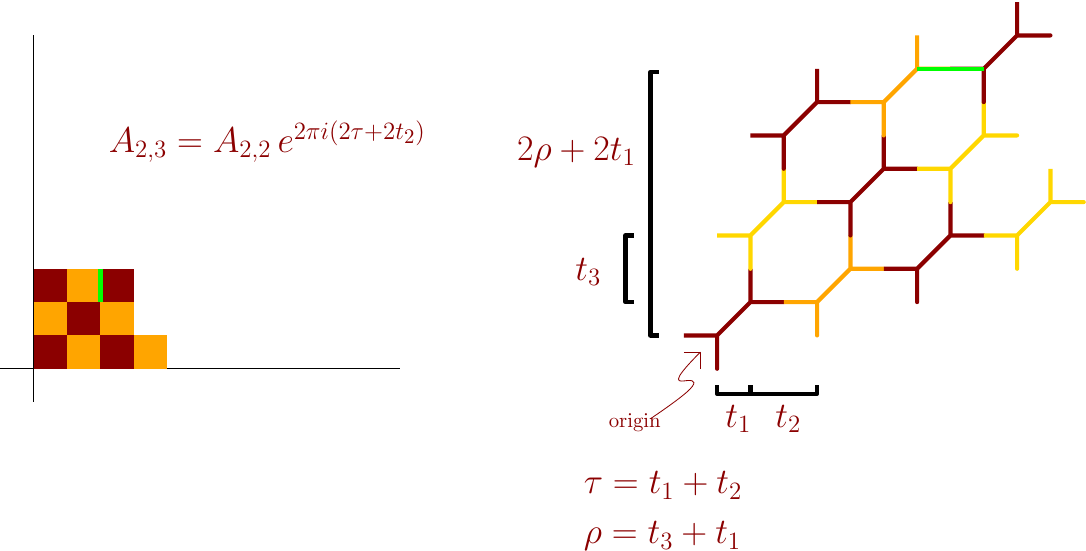}}
        \caption{ tessellation of Newton polygons and web diagram of $X_{1,1}$}
        \label{fig:RC}
      \end{figure}

%\subsection{Example: Resolved Conifold}
We start the discussion by giving an example of Resolved Conifold.
In this case, the Newton polygon is shown in figure (\ref{fig:RC})
and the corresponding mirror curve is given by,
\bea
A_{00}+A_{10}X+A_{01}Y+A_{11}XY=0
\eea
Let us choose  the horizontal line in the web corresponding to the points $(0,0)$ and $(0,1)$  in the Newton polygon that goes through the origin so that $\alpha=0$ for this line. This gives
\bea
A_{01}=A_{00}
\eea
Similarly $A_{10}=A_{00}$ and $A_{10}=A_{01}$. The line in the web corresponding to $(0,1),(1,1)$ has the equation $x=T$ where $T$ is the horizontal distance between the two vertices in the web. Note that  the vertical distance is also $T$. Thus we get $A_{11}=A_{01}e^{-t}$ where $\text{Re}(t)=T$.  The mirror curve is then given by
\bea
1+X+Y+e^{2\pi i t^{*}}X\,Y=0
\eea
 where $t^{*}=\frac{i}{2\pi}t=\frac{i}{2\pi}T-\frac{\mbox{Im}(t)}{2\pi}$ so that $\mbox{Im}(t^{*})>0$.
 %This is the notation that we will use in the rest of the section.
\subsection{Mirror curve dual to $X_{1,1}$}
Recall that in the mirror construction the Riemann surface $\Sigma$ is a part of the mirror CY threefold.  For $6D$ theories the corresponding toric webs have no semi-infinite lines and hence no punctures. The periodicity of the web is taken into account by including all of its images under the periodic shift. Note  that after the vertical and horizontal periodic identifications the toric diagram becomes non-planar.\\
In this case the mirror curve is given by,
\bea
\sum_{(k,\ell)\in \mathbb{Z}^{2}}A_{k,\ell}X^{k}Y^{\ell}=0\,.
\eea
Let's take the origin of the web to be the vertex of the web corresponding to the triangle coordinatized by $(0,0),(1,0),(0,1)$.
With this choice the equation of the horizontal line in the web corresponding to $(k,\ell)$ and $(k,\ell+1)$ is given by
\bea
y=\ell (t_{1}+t_{3})+k\,t_{1}
\eea
where $\tau$ is the periodicity of the web in the vertical direction and $t_{1}$ is the horizontal distance between two consecutive vertices on the diagonal in the web given in figure (\ref{fig:RC}). This gives
\bea\label{eqh}
A_{k,\ell+1}=A_{k,\ell}e^{2\pi i (\ell \tau+k\,z)} \implies A_{k,\ell+1}=A_{k,0}e^{2\pi i (\tau \frac{\ell(\ell+1)}{2}+(\ell+1)k\,z)}
\eea
where $\mbox{Im}(\tau)=\frac{t_{1}+t_{3}}{2\pi}$ and $\mbox{Im}(z)=\frac{t_{1}}{2\pi}$.
The equation of the line in the web corresponding to $(k,\ell),(k+1,\ell)$ is given by $x=k (t_{1}+t_{2})+\ell t_{1}$ where $\rho$ is the periodicity of the web in the horizontal direction. We thus get
\bea\label{eqv}
A_{k+1,\ell}=A_{k,\ell}e^{2\pi i (k \rho+\ell z)}\implies A_{k+1,\ell}=A_{0,\ell}e^{2\pi i (\rho \frac{k(k+1)}{2}+(k+1)\ell z)}
\eea
From Eq.(\ref{eqh}) and Eq.(\ref{eqv}) it follows that
\bea
A_{k,\ell}=A_{0,0}e^{2\pi i (\frac{\ell(\ell-1)}{2}\tau+\frac{k(k-1)}{2}\rho+\ell k z)}
\eea
Using the coefficients  the  mirror curve becomes
\bea
\sum_{k,\ell\in\mathbb{Z}}e^{2\pi i (\frac{\ell(\ell-1)}{2}\tau+\frac{k(k-1)}{2}\rho+\ell k z)}X^{k}Y^{\ell}=0
\eea
If we define the genus two theta function by
\bea
\Theta\Big(\Omega(\rho,z,\tau)|(u,v)\Big)=\sum_{k,\ell}\mbox{exp}\Big(2\pi i Q(k,\ell)/2\Big)X^k Y^{\ell}
\eea
where the period matrix $\Omega(\rho,z,\tau)$ and the quadratic form $Q(k,.)$ are given by
\bea
\Omega(\rho,z,\tau):=\left(
         \begin{array}{cc}
           \rho & z \\
           z & \tau \\
         \end{array}
       \right)\,,\,\,\,\,\,Q(k,\ell):=(k\,\,\ell)\Omega(\rho,z,\tau)\left(
                                                         \begin{array}{c}
                                                           k \\
                                                           \ell \\
                                                         \end{array}
                                                       \right)
\eea
the mirror curve can be written as
\bea\label{eq:curve11}
\Theta\Big(\Omega(\rho,z,\tau)|(u,v)\Big)=0
\eea
It is interesting to note \cite{Hollowood_2008,Haghighat_2019} that under the following identifications
\bea
X\to Xe^{2\pi i\tau},&&\quad Y\to Ye^{z}\nonumber\\
Y\to Y^{2\pi i\rho},  &&\quad X\to Xe^z
\eea
the theta function transforms covariantly and the curve  (\ref{eq:curve11}) remains invariant
\footnote{Recall \cite{Alvarez-Gaume_166751}  the theta function with characteristics given by
\bea
\Theta  \left[ {\begin{array}{cc}
 \bf{a} \\
  \bf{b} \\
  \end{array} } \right](\bf{z}|\Omega)&=&\sum_{\bf{n}\in\mathbb{Z}^g} exp\big(i\pi(\bf{n}+\bf{a}).\Omega.(\bf{n}+\bf{a})+2\pi i(\bf{n}+\bf{a}).(\bf{z}+\bf{b}) \big)\nonumber
\eea
satisfies the following identities 
under the shifts of $\bf{z}$ by lattice L$_{\Omega}$ and $\bf{a},\bf{b}\in \mathbb{R}^g$
\bea
\Theta  \left[ {\begin{array}{cc}
 \bf{a} \\
  \bf{b} \\
  \end{array} } \right](\bf{z}+\Omega\bf{n}+\bf{m}|\Omega)&=&e^{-i\pi \bf{n}.\Omega.\bf{n}-2\pi i\bf{n}.(\bf{z}+\bf{b})+2\pi i \bf{a}.\bf{m}}\Theta  \left[ {\begin{array}{cc}
 \bf{a} \\
  \bf{b} \\
  \end{array} } \right](\bf{z}|\Omega)\nonumber\\
  \Theta  \left[ {\begin{array}{cc}
 \bf{a}+\bf{n} \\
  \bf{b}+\bf{m} \\
  \end{array} } \right](\bf{z}|\Omega)&=&e^{2\pi i \bf{a}.\bf{m}}\Theta  \left[ {\begin{array}{cc}
 \bf{a} \\
  \bf{b} \\
  \end{array} } \right](\bf{z}|\Omega)\nonumber
\eea
}. 
Note that in the limit $z\to 0$ the left side is factorized into the product of genus one theta functions
\bea
\big(\sum_{k,\in\mathbb{Z}}e^{2\pi i (\frac{k(k-1)}{2}\rho)}X^{k}\big)\big(\sum_{\ell\in\mathbb{Z}}
e^{2\pi i (\frac{\ell(\ell-1)}{2}\tau)}Y^{\ell}\big)=0
\eea

\subsection{Mirror curve dual to $X_{1,2}$}
\begin{figure}[!htb]
        \center{\includegraphics[width=6in]{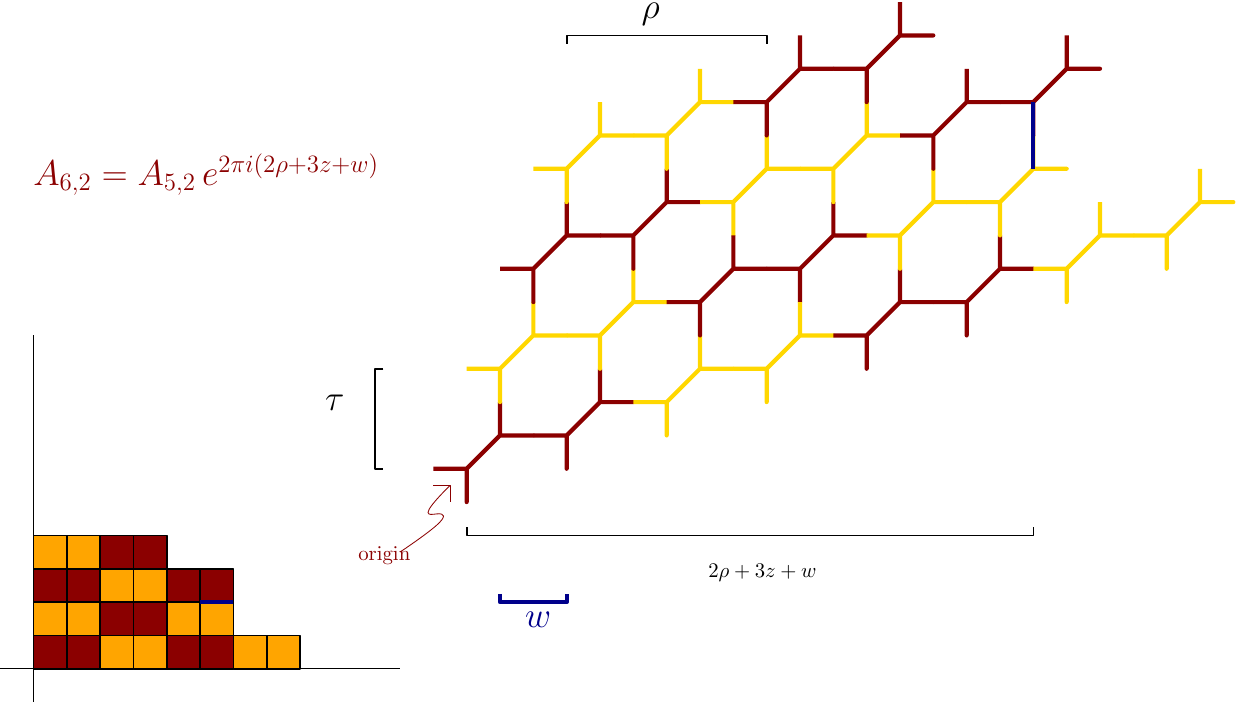}}
        \caption{\label{fig:X12} tessellation of Newton polygons and web diagram of $X_{1,2}$}
      \end{figure}

Consider the periodic Newton polygon with vertices $(0,0),(1,0),(2,0),(2,1),(1,1),(0,1)$ as shown in figure (\ref{fig:X12}). The mirror curve is given by
\bea
\sum_{k,\ell\in \mathbb{Z}}B_{k\ell}\,X^{k}Y^{\ell}=0
\eea
where the coefficients $B_{k,\ell}$ can be determined in the same way as for the genus two case and are functions of the four K\"ahler parameters $(\tau,\rho,z,w)$.
They are related to each other as follows:
\bea
B_{2k+2,\ell}=B_{2k+1,\ell}e^{2\pi i  (k\rho+(\ell+1)z+w)}\,,\,\,\,B_{2k+1}=B_{2k,\ell}e^{2\pi i(k\rho+\ell z)},\,\,\,
B_{k,\ell+1}=B_{k,\ell}e^{2\pi i(\ell\,\tau+k\,z)}
\eea

These recursive relations have the following solution:
\bea\nonumber
B_{2k,\ell}&=&\mbox{exp}\Big[2\pi i\Big(k(k-1)\rho+\frac{\ell(\ell-1)}{2}\tau+2k\ell z+kz+kw\Big)\Big]\\\nonumber
B_{2k+1,\ell}&=&\mbox{exp}\Big[2\pi i \Big(k^2\rho+\frac{\ell(\ell-1)}{2}\tau+(2k+1)\ell z+k(z+w)\Big)\Big]
\eea
Then the mirror curve is given by
\bea
\Theta\Big(\Omega(2\rho,2z,\tau)|(2u-\rho+z+w,v-\tau)\Big)+e^{2\pi i u}\Theta\Big(\Omega(2\rho,2z,\tau)|(2u+z+w,v-\tau+z)\Big)=0\nonumber\\
\eea
To see the factorisation we can write the last expression explicitly as
\bea
\sum_{k,\ell\in \mathbb{Z}}\bigg(\mbox{exp}\Big[2\pi i\Big(&&k(k-1)\rho+\frac{\ell(\ell-1)}{2}\tau+2k\ell z+kz+kw\Big)\Big]X^{2k}Y^{\ell}\nonumber\\
+&&\mbox{exp}\Big[2\pi i \Big(k^2\rho+\frac{\ell(\ell-1)}{2}\tau+(2k+1)\ell z+k(z+w)\Big)\Big]X^{2k+1}Y^{\ell}\bigg)=0
\eea
 It is easy to see that In the limit $z\to 0$ we get the factorized form
%\bea
%&\sum_{k,\ell\in \mathbb{Z}}\bigg(\mbox{exp}\Big[2\pi i\Big(k(k-1)\rho+\frac{\ell(\ell-1)}{2}\tau+kw\Big)\Big]X^{2k}Y^{\ell}
%+\mbox{exp}\Big[2\pi i \Big(k^2\rho+\frac{\ell(\ell-1)}{2}\tau+kw\Big)\Big]X^{2k+1}Y^{\ell}\bigg)=0\nonumber\\
%\eea
\bea
\bigg(\sum_{\ell\in \mathbb{Z}}\mbox{exp}\Big[2\pi i\Big(\frac{\ell(\ell-1)}{2}\tau\Big)\Big]Y^{\ell}\bigg)
\bigg(\sum_{k\in \mathbb{Z}}X^{2k}\bigg(&\mbox{exp}\Big[2\pi i\Big(k(k-1)\rho+kw\Big)\Big]
+\mbox{exp}\Big[2\pi i \Big(k^2\rho+kw \Big)\Big]X\bigg)\bigg)=0\nonumber\\
\eea
%Recall that in the mirror construction this Riemann surface $\Sigma$ is a part of the mirror CY threefold.  For $6D$ theories the corresponding toric webs have no semi-infinite lines and hence no punctures. The periodicity of the web is taken into account by including all of its images under the periodic shift.
\subsection{Mirror curve dual to $X_{N,M}$}\label{secXMN}
\begin{figure}[hbt!]
        \center{\includegraphics[width=6in]{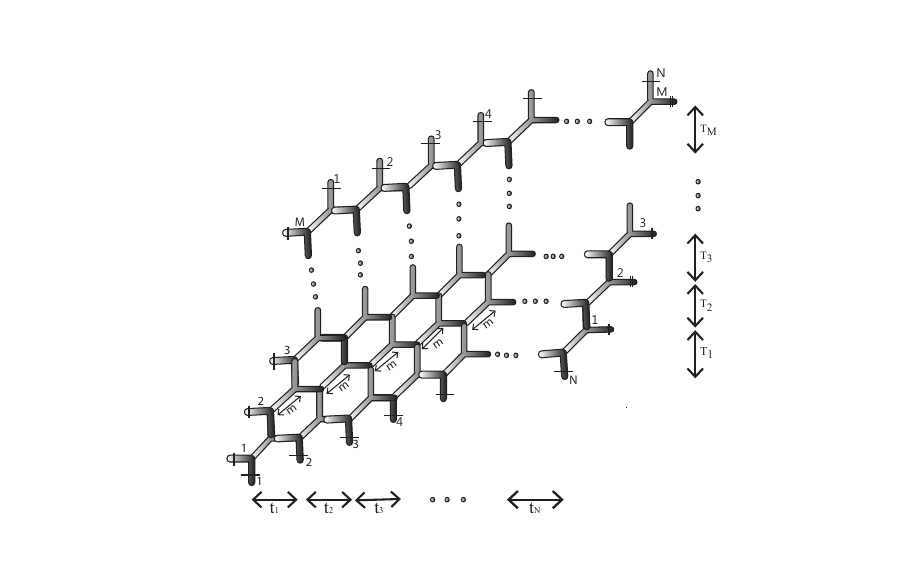}}
    %    \caption{\label{fig:X12} tessellation of Newton polygons and web diagram of $X_{1,2}$}
    \caption{Web diagram of $X_{N,M}$. $t_i\in \{t_1,...,t_N\}$ denotes the distance between $i$-th and $i+1$-th vertical lines and $T_i\in \{T_1,...,T_M\}$ denotes the distance between $i$-th and $i+1$-th horizotntal lines. $\bf{m}_{a,b}$ parametrize the diagonal $\mathbb{P}^1$s. }
\label{Fig:WebToric2}
      \end{figure}
Consider the $(N,M)$ web shown in figure (\ref{Fig:WebToric2}). The K\"ahler class $\omega$ of $X_{N,M}$ is parameterized by $(m_{\alpha,\beta},\tau,\rho,{\bf T},{\bf t})=(m_{\alpha,\beta},\tau,\rho,m,T_{1},T_{2},\cdots,T_{M-1},t_{1},t_{2},\cdots,t_{N-1})$ with $\tau=\sum_{i=1}^{M}T_{i}$ and $\rho=\sum_{j=1}^{N}t_{j}$.
% In the partition function $\mathcal{Z}_{N,M}$ the K\"ahler parameters are classical whereas in the mirror curve the K\"ahler parameter have to be classical.
For arbitrary $(N,M)$ values the factorisation properties of the mirror curve will in general be affected by the quantum corrections. The quantum corrected K\"ahler parameters are the solutions of the Picard-Fuchs equations \cite{Iqbal:2001kk}. After getting quantum corrections  various K\"ahler parameters are mixed non-trivially and that renders the factorisation non-trivial as compared to the classical case discussed here. 
%The Picard Fuch's differential operators can conveniently be written in terms of the sigma model charges $l_i^{(\alpha)}$ and the complex structure parameters $a_i$  as
%\bea
%\mathcal{D}_{\alpha}:=\prod_{l_i^{(\alpha)}>0}(\frac{\partial}{\partial a_i})^{l_i^{(\alpha)}}-\prod_{l_i^{(\alpha)}<0}(\frac{\partial}{\partial a_i})^{-l_i^{(\alpha)}}
%\eea
%To illustrate the point consider the case of local CY $F_2$ for which the classical K\"ahler parameters $(z_1,z_2,z_3,z_4)$ and quantum corrected K\"ahler parameters $(q_1,q_2,q_3,q_4)$ satisfy the following relations
%\bea
%z_1&=&q_1\sqrt{\frac{z_2}{q_2}}(\frac{z_3}{q_3})^2\nonumber\\
%z_4&=&q_4\frac{z_2}{q_2}\frac{z_3}{q_3}
%\eea
%This shows that when expressed in terms of quantum corrected parameters, the mirror curves will not have similar degeneration properties.
 \\The mirror curve is given by a sum over the monomials associated with the Newton polygon. In this case the Newton polygon tiles the plane 
\bea
H_{N,M}(X,Y):=\sum_{(i,j)\in \mathbb{Z}^2}A_{i,j}X^{i}\,Y^{j}\,.
\eea
 \footnote{ The notation $H_{N,M}$ should not be confused with  $\mbox{H}$ which denotes the instanton moduli space in the introduction.}
The coefficients $A_{i,j}$ depend on the length of the various line segments in the web which are the K\"ahler parameters of the corresponding Calabi-Yau threefolds. As discussed before the neighbouring pair of points in the Newton polygon connected by a line give a relation between the associated coefficients $A_{i,j}$,
\bea
\frac{A_{i,k+1}}{A_{i,k}}=e^{\sum_{j=1}^{k-1}T_{j}+\sum_{\alpha=0}^{i-1}m_{\alpha,k}}\\\nonumber
\frac{A_{i+1,k}}{A_{i,k}}=e^{\sum_{j=1}^{i-1}t_{j}+\sum_{\alpha=0}^{k-1}m_{i,\alpha}}
\eea
\bea
A_{i+1,k+1}&=&A_{i+1,1}e^{T_{1}+(T_{1}+T_{2})+(T_{1}+T_{2}+T_{3})+\cdots+(T_{1}+\cdots+T_{k-1})+\sum_{\beta=1}^{k}\sum_{\alpha=0}^{i}m_{\alpha,\beta}}\\\nonumber
&=&A_{i+1,1}e^{\sum_{\gamma=1}^{k-1}(k-\gamma)T_{\gamma}+\sum_{\beta=1}^{k}\sum_{\alpha=0}^{i}m_{\alpha,\beta}}\\\nonumber
&=&A_{0,1}e^{t_{1}+(t_{1}+t_{2})+\cdots+(t_{1}+t_{2}+\cdots+t_{i-1})}
e^{\sum_{\gamma=1}^{k-1}(k-\gamma)T_{\gamma}+\sum_{\beta=0}^{k}\sum_{\alpha=0}^{i}m_{\alpha,\beta}}\\\nonumber
\eea
where in the  web diagram of $X_{N,M}$, $t_i\in \{t_1,...,t_N\}$ denotes the distance between $i$-th and $i+1$-th vertical lines and $T_i\in \{T_1,...,T_M\}$ denotes the distance between $i$-th and $i+1$-th horizotntal lines and $\bf{m}_{a,b}$ parametrize the diagonal finite line segments representing $\mathbb{P}^1$s.\\
Using $A_{0,1}=A_{0,0}=1$ we get the following solution
\bea
A_{i+1,k+1}&=&e^{\sum_{\gamma=1}^{i-1}(i-\gamma)t_{\gamma}+\sum_{\gamma=1}^{k-1}(k-\gamma)T_{\gamma}+\sum_{\beta=0}^{k}\sum_{\alpha=0}^{i}m_{\alpha,\beta}}\\\nonumber
\eea
Thus the curve is given by
\bea
H_{N,M}(X,Y)&=&\sum_{(i,k)\in \mathbb{Z}^2}A_{i+1,k+1}X^{i+1}Y^{k+1}\\\nonumber
&=&\sum_{i=0,k=0}^{N-1,M-1}W_{i,k}(X,Y)\\\nonumber
W_{i,k}(X,Y)&=&\sum_{(a,b)\in \mathbb{Z}^2}A_{Na+i+1,Mb+k+1}X^{Na+i+1}Y^{Mb+k+1}
\eea
\bea
A_{Na+i+1,Mb+k+1}=e^{\sum_{\gamma=1}^{Na+i-1}(Na+i-\gamma)t_{\gamma}+\sum_{\gamma=1}^{Mb+k-1}(Mb+k-\gamma)T_{\gamma}+
\sum_{\beta=0}^{Mb+k}\sum_{\alpha=0}^{Na+i}m_{\alpha,\beta}}\\\nonumber
\eea
Using the identifications
\bea
t_{\gamma}&=&t_{\gamma'}\,\,\,\,\mbox{if}\,\,\,\,\,\gamma\equiv\gamma'\,(\mbox{mod}\,N)\\\nonumber
T_{\gamma}&=&T_{\gamma'}\,\,\,\,\mbox{if}\,\,\,\,\,\gamma\equiv\gamma'\,(\mbox{mod}\,M)\\\nonumber
m_{\alpha_{1},\beta_{1}}&=&m_{\alpha_{2},\beta_{2}}\,\,\,\,\,\mbox{if}\,\,\,\,\,\,\,\alpha_{1}\equiv\alpha_{2}\,(\mbox{mod}\,N)\,\,\,\mbox{and}\,\,\,\,\beta_{1}\equiv\beta_{2}\,(\mbox{mod}\,M)
\eea
we get
\bea
\sum_{\gamma=1}^{Na+i-1}(Na+i-\gamma)t_{\gamma}&=&\sum_{\gamma=1}^{N}(Na+i-\gamma)t_{\gamma}+\sum_{\gamma=N+1}^{2N}(Na+i-\gamma)t_{\gamma}+\cdots\\\nonumber
&&+\sum_{\gamma=N(a-1)+1}^{Na}(Na+i-\gamma)t_{\gamma}+\sum_{\gamma=Na+1}^{Na+i-1}(Na+i-\gamma)t_{\gamma}\\\nonumber
&=&\sum_{\gamma=1}^{N}\Big[(Na+i-\gamma)+(N(a-1)+i-\gamma)+(N(a-2)+i-\gamma)+\cdots+\\\nonumber
&&(N+i-\gamma)\Big]t_{\gamma}+\sum_{\gamma=1}^{i-1}(i-\gamma)t_{\gamma}\\\nonumber
&=&\sum_{\gamma=1}^{N}\Big[N\tfrac{a(a+1)}{2}+a(i-\gamma)\Big]t_{\gamma}+\sum_{\gamma=1}^{i-1}(i-\gamma)t_{\gamma}\\\nonumber
&=&\Big[N\tfrac{a(a+1)}{2}+ai\Big]\tau-\sum_{\gamma=1}^{N}\gamma\,t_{\gamma}+\sum_{\gamma=1}^{i-1}(i-\gamma)t_{\gamma}
\eea
Similarly
\bea
\sum_{\gamma=1}^{Mb+k-1}(Mb+k-\gamma)T_{\gamma}=\Big[M\tfrac{b(b+1)}{2}+bk\Big]\rho-\sum_{\gamma=1}^{M}\gamma\,T_{\gamma}+\sum_{\gamma=1}^{k-1}(k-\gamma)T_{\gamma}
\eea
\bea
\sum_{\beta=0}^{Mb+k}\sum_{\alpha=0}^{Na+i}m_{\alpha,\beta}&=&\sum_{\beta=0}^{Mb+k}\Big[\sum_{\alpha=0}^{N-1}m_{\alpha,\beta}+\sum_{\alpha=N}^{2N-1}m_{\alpha,\beta}+\cdots+
\sum_{\alpha=N(a-1)}^{Na-1}m_{\alpha,\beta}+\sum_{\alpha=Na}^{Na+i}m_{\alpha,\beta}\Big]\\\nonumber
&=&\sum_{\beta=0}^{Mb+k}\Big[a\sum_{\alpha=0}^{N-1}m_{\alpha,\beta}+\sum_{\alpha=0}^{i}m_{\alpha,\beta}\Big]\\\nonumber
&=&a\sum_{\alpha=0}^{N-1}\Big[b\sum_{\beta=0}^{M-1}m_{\alpha,\beta}+\sum_{\beta=0}^{k}m_{\alpha,\beta}\Big]+\sum_{\alpha=0}^{i}
\Big[b\sum_{\beta=0}^{M-1}m_{\alpha,\beta}+\sum_{\beta=0}^{k}m_{\alpha,\beta}\Big]\\\nonumber
&=&ab\sum_{\alpha=0}^{N-1}\sum_{\beta=0}^{M-1}m_{\alpha,\beta}+a\sum_{\alpha=0}^{N-1}\sum_{\beta=0}^{k}m_{\alpha,\beta}+b\sum_{\alpha=0}^{i}\sum_{\beta=0}^{M-1}m_{\alpha,\beta}+
\sum_{\alpha=0}^{i}\sum_{\beta=0}^{k}m_{\alpha,\beta}
\eea
Since $\sum_{\alpha=0}^{N-1}m_{\alpha,\beta}$ is independent of $\beta$ by Lemma 5.4 of \cite{Kanazawa:2016tnt}
\footnote{Note that  $m_{\alpha,\beta}$ is denoted as $C^1_{(a,b)}$ in \cite{Kanazawa:2016tnt}} 
therefore
\bea
\sum_{\beta=0}^{Mb+k}\sum_{\alpha=0}^{Na+i}m_{\alpha,\beta}&=&(ab+\tfrac{a(k+1)}{M}+\tfrac{b(i+1)}{N})
\sum_{\alpha=0}^{N-1}\sum_{\beta=0}^{M-1}m_{\alpha,\beta}+
\sum_{\alpha=0}^{i}\sum_{\beta=0}^{k}m_{\alpha,\beta}\\\nonumber
&=&(ab+\tfrac{a(k+1)}{M}+\tfrac{b(i+1)}{N}){\bf m}+{\bf m}^{i,k}
\eea
\bea
&&\sum_{\gamma=1}^{Na+i-1}(Na+i-\gamma)t_{\gamma}+\sum_{\gamma=1}^{Mb+k-1}(Mb+k-\gamma)T_{\gamma}+
\sum_{\beta=0}^{Mb+k}\sum_{\alpha=0}^{Na+i}m_{\alpha,\beta}+\\\nonumber
&&z_{1}(Na+i+1)+z_{2}(Mb+k+1)=\\\nonumber
&&\Big[N\tfrac{a(a+1)}{2}+ai\Big]\tau-\sum_{\gamma=1}^{N}\gamma\,t_{\gamma}+\sum_{\gamma=1}^{i-1}(i-\gamma)t_{\gamma}+
\Big[M\tfrac{b(b+1)}{2}+bk\Big]\rho-\sum_{\gamma=1}^{M}\gamma\,T_{\gamma}+\sum_{\gamma=1}^{k-1}(k-\gamma)T_{\gamma}+\\\nonumber
&&(ab+\tfrac{a(k+1)}{M}+\tfrac{b(i+1)}{N}){\bf m}+{\bf m}^{i,k}+z_{1}(Na+i+1)+z_{2}(Mb+k+1)\\\nonumber
&=&G^{i,k}_{N,M}({\bf t},{\bf T},{\bf m})+\tfrac{1}{2}(a+\tfrac{i+1}{N},b+\tfrac{k+1}{M})\begin{pmatrix}N\tau & {\bf m}\\ {\bf m}& M\rho\end{pmatrix}\begin{pmatrix}a+\tfrac{i+1}{N}\\ b+\tfrac{k+1}{M}\end{pmatrix}+a\tau(\tfrac{N}{2}-1)+b\rho(\tfrac{M}{2}-1)\\\nonumber
&&-\tfrac{(i+1)(k+1)}{MN}{\bf m}-\tfrac{1}{2}(\tfrac{i+1}{N})^2\,N\tau-\tfrac{1}{2}(\tfrac{k+1}{M})^2\,M\rho+Nz_{1}(a+\tfrac{i+1}{N})+Mz_{2}(b+\tfrac{k+1}{M})\\\nonumber
&=&G^{i,k}_{N,M}({\bf t},{\bf T},{\bf m})+\tfrac{1}{2}({\bf n}+{\bf u})^{t}\Omega ({\bf n}+{\bf u})+({\bf n}+{\bf u})\cdot (\widehat{\bf z}+{\bf v})
\eea
where
\bea
G^{i,k}_{N,M}({\bf t},{\bf T},{\bf m})&=&-\tfrac{(k+1)(M+k-1)}{2M}\rho-\tfrac{(i+1)(N+i-1)}{2N}\tau+
{\bf m}^{i,k}-\sum_{\gamma=1}^{N}\gamma\,t_{\gamma}+\sum_{\gamma=1}^{i-1}(i-\gamma)t_{\gamma}\nonumber\\
&&-\sum_{\gamma=1}^{M}\gamma\,T_{\gamma}+\sum_{\gamma=1}^{k-1}(k-\gamma)T_{\gamma}\\\nonumber
\widehat{\bf z}&=&(Nz_{1},Mz_{2})\\\nonumber
{\bf u}&=&(\tfrac{i+1}{N},\tfrac{k+1}{M})\\\nonumber
{\bf v}&=&(\tau(\tfrac{N}{2}-1),\rho(\tfrac{M}{2}-1))
\eea
We define the genus two theta function as:
\bea\label{eq:Theg2theta}
\Theta_{\vec{u},\vec{v}}(\vec{z},\Omega)=\sum_{\vec{n}\in \mathbb{Z}^2}e^{\frac{1}{2}(\vec{n}+\vec{u})^{t}\Omega(\vec{n}+\vec{u})+(\vec{n}+\vec{u})\cdot(\vec{z}+\vec{v})}
\eea
Then
\bea
W^{i,k}(X,Y)=\sum_{(a,b)\in \mathbb{Z}^2}e^{G^{i,k}_{N,M}}\,\Theta_{\vec{u},\vec{v}}(\vec{z},\Omega)
\eea
The genus of the mirror curve
\bea
\sum_{i=0,k=0}^{N-1,M-1}W_{i,k}(X,Y)=0
\eea
is $MN+1$. The underlying abelian surface has polarisation $(N,M)$ with the period matrix given by $\Omega=\begin{pmatrix}N\tau & {\bf m}\\ {\bf m}& M\rho\end{pmatrix}$. The theta functions form a basis corresponding to this $(N,M)$-polarization of the abelian surface.
% The  mirror curve for general values of $N$ and $M$ cannot be factorized in the limit $z\to 0$.
\subsection{Geometric interpretation of the mirror curve}
An illuminative way to visualise the mirror curve $\Sigma$ is to see it as N copies of the base torus glued together by N-1 branch cuts \cite{Braden:2003gv,Hollowood_2008}. The one cycles, A and B, of the base torus are lifted to a basis of 1-cycles $A_i, B_i, i=1,...,N$ on $\Sigma$. Riemann-Hurwitz theorem is used to compute the genus  of $\Sigma$ and is equal to N+1. The  Riemann-Roch theorem is handy in the computation of the number of moduli of $\Sigma$, which is equal to N in this case. \\
In the case under consideration, the genus N Riemann surface is seen as defined by theta divisor. A general polarised abelian variety $\mathcal{U}$ admits a line bundle $\mathcal{L}$ with $c_1(\mathcal{L})=\omega$ where $\omega$ is a $(1,1)$-form that  is given in terms of the coordinates  $0\le y_i\le1$   by
\bea
\omega=[N dy_1\wedge dy_3+dy_2\wedge dy_4]
\eea
where it is assumed that  the period matrix $\Omega$ of $\Sigma$ is symmetric and $Im(\Omega)>0$. For general abelian variety with polarisation given by $\omega=[N dy_1\wedge dy_3+M dy_2\wedge dy_4]$ the line bundle $\mathcal{L}$ admits $MN$ holomorphic sections. In the case of an abelian surface these sections are given by genus $2$ theta functions
\bea
\Theta  \left[ {\begin{array}{cc}
  \frac{i}{M}& \frac{j}{N} \\
  0 &0 \\
  \end{array} } \right](z|\Omega)\quad 0\le i<M,\quad 0\le j<N.
\eea
A theta divisor is the zero locus of a linear combination of the above set of theta functions
\bea
\sum_i^M\sum_j^N A_{ij}\Theta  \left[ {\begin{array}{cc}
  \frac{i}{M}& \frac{j}{N} \\
  0 &0 \\
  \end{array} } \right](\vec{z}|\Omega)=0
\eea
where $A_{ij}$ denote the moduli of the curve.
This zero locus defines the mirror curve of  genus $MN+1$ and is the Riemann surface $\Sigma$.
For the special case of $M=1$ the mirror curve can be  expressed in the following form
\bea
\sum_{n=0}\frac{1}{n!}(\frac{m}{2\pi i})^n\partial_z^n\theta_1(z|\tau)\partial_x^n h(x)=0
\eea
where $\theta_1$ is the Jacobi theta function and $h(x)=\prod_{j=1}^N\theta_1(x-\xi_j|\rho)$ with $\xi_j$ is the moduli of $\Sigma$. This can be reorganised into the following form
\bea
\Theta_{[\frac{1}{2},...,\frac{1}{2}],[\frac{1}{2},...,\frac{1}{2}]}(z, \frac{N\beta}{2\pi}(x-\xi_i)| \hat{\Omega})=0
\eea
where $\hat{\Omega}$ is the period matrix of the genus MN+1 curve $\hat{\Sigma}$ which is an unbranched cover of a genus 2 curve and in general  is given by \\
\bea
 \hat{\Omega}=
  \left[ {\begin{array}{ccccc}
   \tau & \frac{\beta m_1}{2\pi i} &  \frac{\beta m_2}{2\pi i} &  \frac{\beta m_3}{2\pi i} &...  \frac{\beta m_{MN}}{2\pi i}\\
  \frac{\beta m_1}{2\pi i}  & \rho & 0 & 0 &... 0\\
    \frac{\beta m_2}{2\pi i}  & 0 & \rho & 0 &... 0\\
   . & . & . & .  &\\
    . & . & . & .  &\\
     . & . & . & .  &\\
     \frac{\beta m_{MN}}{2\pi i}    & 0 & 0& 0  &...\rho\\
  \end{array} } \right]
\eea
It is easy to see
from the following  representation of genus $g=MN+1$ theta function
\bea
\Theta \left[ {\begin{array}{c}
  \alpha \\
\beta
  \end{array} } \right](Z|\hat{\Omega})=\sum_{m\in \mathbb{Z}^g}exp\bigg(\pi i (m+\alpha).\hat{\Omega}.(m+\alpha)+2\pi i (Z+\beta).(m+\alpha)  \bigg)
\eea
where $Z,\alpha,\beta,m$ are g-vectors and $\Omega$ is a $g\times g$ matrix with $Im\Omega>0$.\\
%\bea
%\Theta^{g=N+1}=\sum_{i\in \mathbb{Z}_{N+1}}\theta^{g=1}\Theta^{g=N}
%\eea
%%%%%%%%%%%
% In the limit $m_N\to 0$ the genus $N+1$ theta function gets split into the product of  genus $N$ theta function and a Jacobi theta function $(g=1)$ with characteristics.
%%%%%%%%%%%%
%The genus 3 curve mirror  to CYthreefold $X_{1,2}$ is given by
%For instance the  mirror curve for $X_{1,2}$ is given by
%\bea
%\Theta\Big(\Omega(2\rho,2z,\tau)|(2u-\rho+z+w,v-\tau)\Big)+e^{2\pi i u}\Theta\Big(\Omega(2\rho,2z,\tau)|(2u+z+w,v-\tau+z)\Big)=0\nonumber\\
%\eea
%In the limit of $Im(u)\to \infty$ this curve collapses to
%\bea
%\Theta\Big(\Omega(2\rho,2z,\tau)|(2u-\rho+z+w,v-\tau)\Big) =0
%\eea
%which, after performing $Sp(2,\mathbb{Z})$ transformations, is the curve mirror to $X_{1,1}$.
%XXXXXXXXXX
%\subsection{Splitting of higher genus theta functions and M5-branes partition functions}
To study the decomposition of generalised theta function \cite{Marshakov:1999bw}  defined on the Jacobian of a genus $g=M$ curve, we start from
the following Fourier representation
\bea\label{eq:Theta}
\Theta(\Omega|\vec{z})=\sum_{\textbf{m}\in \mathbb{Z}^M}e^{2\pi i \sum_{i=1}^Mm_iz_i+i\pi\sum_{i,j=1}^Mm_i\Omega_{ij}m_j}
\eea
where $\Omega$ is the period matrix and satisfies the following constraints
\bea
\sum_{i=1}^M\Omega_{ij}&=&\tau,\quad
\sum_{j=1}^M\Omega_{ij}=\tau
\eea
This constraint encodes various periodicity properties. In other words we can decompose $\Omega$ as
\bea
\Omega=\frac{\tau}{M}+\Omega^{\prime}
\eea
where $\Omega^{\prime}$ is the traceless part. Now redefine $z_i$ as follows
 \bea
 z_i&=&\frac{z}{M}+z_i^{\prime}\quad  \mbox{such that}\quad
 \sum_{i=1}^M z_i^{\prime}=0
 \eea
Putting back these redefined variables in (\ref{eq:Theta}) we get
\bea\label{eq:thetasplitting}
\Theta(\Omega|\vec{z})&=&\sum_{\textbf{m}\in \mathbb{Z}^M} e^{2\pi i \frac{z}{M}\sum_{i=1}^Mm_i+i\pi \frac{\tau}{M}(\sum_{i=1}^Mm_i)^2+2\pi i \sum_{i=1}^Mm_i z_i^{\prime}+\pi i \sum_{i,j=1}^Mm_i\Omega^{\prime}_{ij}m_i}\nonumber\\ &=&\sum_{i\in\mathbb{Z}_M,s\in \mathbb{Z}} e^{2\pi i(s+\frac{i}{M})z+\pi i\tau M(s+\frac{i}{M})^2}\sum_{\textbf{m}\in\mathbb{Z}^M,\sum_{j=1}^M m_j=i} e^{ i 2\pi  \sum_{p=1}^Mm_p z_p^{\prime}+\pi i \sum_{p,q=1}^Mm_p\Omega^{\prime}_{pq}m_q}\nonumber\\
&=&\sum_{i\in\mathbb{Z}_M}\theta \left[ {\begin{array}{c}
  \frac{i}{M} \\
0
  \end{array} } \right](M\tau| z)\Theta_i(\Omega^{\prime}| \vec{z}^{\prime})
\eea
%Next we use a trick, essentially a redefinition of indices, to write the exponential in a suggestive form. To this end we decompose the set of indices $\textbf{m}$ into two parts. First we impose the constraint that $\sum_{i=1}^M m_i=l$ which effectively reduces
%the set $\{m_1,...,m_M\}$ to $\{m_1,...,m_{M-1}\}$. Secondly we perform a sum over $l$.
%\bea
%\Theta(\Omega|\vec{z})=\sum_{l\in \mathbb{Z}} e^{2\pi i\frac{l}{M}z+\pi i\frac{l^2}{M}\tau}\sum_{\textbf{m}\in\mathbb{Z}^M;\sum_{i=1}^M m_i=l} e^{ i2\pi i \sum_{i=1}^Mm_i z_i^{\prime}+\pi i \sum_{i,j=1}^Mm_i\Omega^{\prime}_{ij}m_j}
%\eea\label{eq:splittheta}
%To be able to write the first summation as a Jocobi theta function with characteristics, we make another redefinition $l=Ms+i$ where $s\in\mathbb{Z}$ and $i\in \mathbb{Z}_M$, resulting in
%\bea\label{eq:thetasplitting}
%\Theta(\Omega |\vec{z})   &=&\sum_{i\in\mathbb{Z}_M,s\in \mathbb{Z}} e^{2\pi i(s+\frac{i}{M})z+\pi i\tau M(s+\frac{i}{M})^2}\sum_{\textbf{m}\in\mathbb{Z}^M,\sum_{j=1}^M m_j=i} e^{ i 2\pi  \sum_{p=1}^Mm_p z_p^{\prime}+\pi i \sum_{p,q=1}^Mm_p\Omega^{\prime}_{pq}m_q}\nonumber\\
%&=&\sum_{i\in\mathbb{Z}_M}\theta \left[ {\begin{array}{c}
  %\frac{i}{M} \\
%0
 % \end{array} } \right](M\tau| z)\Theta_i(\Omega^{\prime}| \vec{z}^{\prime})
%\eea
where $\Theta_i$ is  the second summation factor in the first  line of (\ref{eq:thetasplitting}).\\
%The splitting of theta functions has important consequences for M5-branes partition function. On general grounds \cite{Haghighat_2019,Marshakov_2000} the partition functions of M5-branes on a six-manifold $X$  are actually sections of a line bundle $L$ over the intermediate Jacobian $J_X=H^3(X,\mathbb{R})/H^3(W,\mathbb{Z})$. The intermediate Jacobian for our CY threefold is an abelian surface. The line bundle $L$ is uniquely specified by its first Chern class $c_1(L)=\omega$, where $\omega\in H^2(J,\mathbb{Z})$ gives the principal polarisation. For the case at hand we have $M$ M5 branes that are probing the transverse space $\mathbb{S}^1\times \mathbb{C}^2/\mathbb{Z}_N$. The partition function of this theory correspond to sections of a line bundle $L$ of polarisation  $(M,N)$ and readily given by eq.(\ref{eq:Theg2theta})
%\bea\label{eq:M5theta}
%\Theta_{\vec{u},\vec{v}}(\vec{z},\Omega)=\sum_{\vec{n}\in \mathbb{Z}^2}e^{\frac{1}{2}(\vec{n}+\vec{u})^{t}\Omega(\vec{n}+\vec{u})+(\vec{n}+\vec{u})\cdot(\vec{z}+\vec{v})}
%\eea
%and there are $MN$ of them.
%The eq.(\ref{eq:thetasplitting}) then shows that a genus $g$ theta function splits into a product of  genus $g-1$ theta function and an ordinary theta function. Therefore same must be true of the theta functions (\ref{eq:M5theta}) that describe M5-branes partition functions.
\section{Degenerations and their Effect on the Partition Function}
The partition function of the CY threefold  $X_{N,M}$ is given by \cite{Hohenegger:2013ala}
\bea\label{Z(N,M)}
\mathcal{Z}_{(N,M)}(\tau, \rho, \epsilon_{1,2}, m_a=m, t)&&=\sum_{\alpha_{a}^{i}}\,\prod_{i=1}^{N}Q_{i}^{|\alpha^{(i)}|}\,\prod_{i=1}^{N}\prod_{a=1}^{M}\,\frac{\vartheta_{\alpha_{a}^{i+1}\alpha_{a}^{i}}(m)}{\vartheta_{\alpha_{a}^{i}\alpha_{a}^{i}}(\epsilon_{+})}\nonumber\\
&&\prod_{1\leq a<b\leq M}\prod_{i=1}^{N}\frac{\vartheta_{\alpha_{a}^{i}\alpha_{b}^{i+1}}(t_{ab}-m)\vartheta_{\alpha_{a}^{i+1}\alpha_{b}^{i}}(t_{ab}+m)}{\vartheta_{\alpha_{a}^{i}\alpha_{b}^{i}}(t_{ab}-\epsilon_{+})\vartheta_{\alpha_{a}^{i}\alpha_{b}^{i}}(t_{ab}+\epsilon_{+})}
\eea
where the sum is over  $N$ partitions of  $\alpha^{(a)}=\{\alpha_{1}^{(a)}, \alpha_{2}^{(a)},...,\alpha_{N}^{(a)}\}$ and $\alpha_{a}^{(1)}\equiv \alpha_{a}^{(N+1)}$, 
$Q_i= e^{b_{i+1}-b_i}$,
$t_{ab}=t_{a,a+1}+t_{a+1,a+2}+\cdots +t_{a+b-(a+1),b}$,  $b_{i+1}-b_i$ is the distance between vertical lines (or M5 branes) and moreover the factorisation degeneration takes place when all the mass parameters $m_a$ are taken equal to $m$.
%\subsection*{
The expressions of partition functions after degeneration becomes particularly simple at the special point in the K\"ahler moduli space where $Q_i:=Q:=e^{2\pi i \tau}$ and in the unrefined limit of the $\Omega$-background parameters $\epsilon_1=-\epsilon_2=\epsilon$.

%}
 We define
\bea
|\alpha^{(a)}|=\sum_{b=1}^N|\alpha_{b}^{(a)}|
\eea
where $|\alpha^{(a)}|$ is the size of the partition $\alpha^{(a)}$  which is the sum of the parts of partition.
%and $t_{ab}$ is the distance between M2 branes, $t_{ab}=m$, $\epsilon_+=\frac{\epsilon_1+\epsilon_2}{2}.$
%
To study the degeneration of partition function we have to study the $x\to 0$ limit of $\vartheta_{\mu\nu}(x)$.
For two integer partitions $\mu$ and $\nu$, theta function $\vartheta_{\mu\nu}$ in the above partition function \eqref{Z(N,M)} is defined as
\bea
&&\vartheta_{\mu\nu}(x)=\prod_{(i,j)\in \mu}\vartheta(\rho, e^{-x}\,t^{-\nu^{t}_{j}+i-\tfrac{1}{2}}\,q^{-\mu_{i}+j-\tfrac{1}{2}})\,\prod_{(i,j)\in \nu}\vartheta(\rho, e^{-x}\,t^{\mu^{t}_{j}-i+\tfrac{1}{2}}\,q^{\nu_{i}-j+\tfrac{1}{2}})\,
\eea
Here $t=e^{-i\epsilon_2}, q=e^{i\epsilon_1}$, $\nu^{t}$ represents the transpose of the partition $\nu$ and product $\prod\limits_{(i,j)\in \nu}$ means that the product is over all the boxes of the Young diagram corresponding to the partition $\nu$ having length $ \ell(\nu)$
$$(i,j)\in \nu\, ,\quad \textrm{implies that}  \quad 1\leq i \leq \ell(\nu),\quad 1 \leq j \leq \nu_i.$$
The Jacobi theta function $\vartheta(\rho,y)$ for $y=e^{2\pi i z}$ is defined as
\bea\nonumber
\vartheta(\rho, y)= (y^{\tfrac{1}{2}}-y^{-\tfrac{1}{2}})\prod_{k=1}^{\infty}(1-y\,e^{2\pi i k\rho})(1-y^{-1}\,e^{-2\pi i k\rho})
\eea
For $x=0$ and in unrefined case
\bea
\vartheta_{\mu\nu}(0)&&=-\prod_{(i,j)\in \mu}\vartheta(\rho, \nu^{t}_{j}-i+\mu_{i}-j+1)\,\prod_{(i,j)\in \nu}\vartheta(\rho, \mu^{t}_{j}-i+\nu_{i}-j+1)\nonumber\\
&&=-\prod_{(i,j)\in \mu}\vartheta(\rho, h_{\mu}(i,j) +\nu^{t}_{j}-\mu^{t}_{j})\,\prod_{(i,j)\in \nu}\vartheta(\rho, h_{\nu}(i,j) +\mu^{t}_{j}-\nu^{t}_{j})\,
\eea
where $h_{\mu}(i,j)=\mu_{i}+\mu_{j}^{t}-i-j+1$ is the hook length of the partition $\mu$. Since, the Jacobi theta function $\vartheta(\rho, z)$ is an odd function w.r.t. $z$ i.e., $\vartheta(\rho, 0)=0$, therefore $\vartheta_{\mu\nu}(0)=0$ if $h_{\mu}(i,j) +\nu^{t}_{j}-\mu^{t}_{j}=0$. Since $h_{\mu}(i,j)\neq 0$, therefore $\nu^{t}_{j}\neq\mu^{t}_{j}$. 
If $\mu=\nu$ then
\bea
\vartheta_{\mu\mu}(0)=\prod_{(i,j)\in \mu}\vartheta(\rho, h_{\mu}(i,j))^2\nonumber
\eea
$h_{\mu}(i,j)$ is non zero therefore $\vartheta_{\mu\mu}(0)\neq0.$
In other words $\mu \neq \nu$ implies $\vartheta_{\mu\nu}(0)=0$ i.e.
either $h_{\mu}(i,j) +\nu^{t}_{j}-\mu^{t}_{j}=0$ or $h_{\nu}(i,j) +\mu^{t}_{j}-\nu^{t}_{j}=0.$
Because $h_{\mu}(i,j) \neq 0$ therefore $\nu^{t}_{j}\neq\mu^{t}_{j}$.
We thus arrive at the useful  property of $\vartheta_{\mu\nu}(x)$ at $x=0$ given by:
\bea\label{Id1}
\vartheta_{\mu\nu}(0)=\delta_{\mu\,\nu}\,\prod_{(i,j)\in \mu}\vartheta(q^{h_{\mu}(i,j)})\vartheta(q^{-h_{\mu}(i,j)})
\eea
where $\delta_{\mu\,\nu}$ is the kronecker delta function and $h_{\mu}(i,j)=\mu_{i}+\mu_{j}^{t}-i-j+1$ is the hook length of the partition $\mu$. This identity is useful for studying different degenerations of the partition functions.
\section{Degeneration 1:Factorization}
This type of degeneration corresponds to taking both the vertical and horizontal `distances' between the 5-branes equal to $m$, which is the K\"ahler parameter corresponding to the exceptional curve or $(1,1)$ brane in the web diagram fig.\ref{Fig:WebToric2}.
\subsection{$(N,M)=(1,2)$ }
We begin by looking at the case of $X_{1,2}$. The unrefined partition function is given by,
\bea\label{Z(1,2)}
\mathcal{Z}_{(1,2)}(\tau,\rho,m,t,\epsilon)&=&\sum_{\alpha_{1,2}}\,Q^{|\alpha_{1}|+|\alpha_{2}|}\,\frac{\vartheta_{\alpha_{1}\alpha_{1}}(m)\vartheta_{\alpha_{2}\alpha_{2}}(m)}{\vartheta_{\alpha_{1}\alpha_{1}}(0)\vartheta_{\alpha_{2}\alpha_{2}}(0)}
\frac{\vartheta_{\alpha_{1}\alpha_{2}}(t^{-}_{m})\vartheta_{\alpha_{1}\alpha_{2}}(t^{+}_{m})}{\vartheta_{\alpha_{1}\alpha_{2}}(t)^2}
\eea
Here, $t^{-}_{m}=t-m$ and $t^{+}_{m}=t+m$. The  partition function $Z_{(1,2)}$  in the limit $t\mapsto m$ reduces  to
\bea
\mathcal{Z}_{(1,2)}(\tau,\rho,m,\epsilon)&=&\sum_{\alpha_{1,2}}\,Q^{|\alpha_{1}|+|\alpha_{2}|}\,\frac{\vartheta_{\alpha_{1}\alpha_{1}}(m)\vartheta_{\alpha_{2}\alpha_{2}}(m)}{\vartheta_{\alpha_{1}\alpha_{1}}(0)\vartheta_{\alpha_{2}\alpha_{2}}(0)}
\frac{\vartheta_{\alpha_{1}\alpha_{2}}(0)\vartheta_{\alpha_{1}\alpha_{2}}(2m)}{\vartheta_{\alpha_{1}\alpha_{2}}(m)^2}\nonumber
\eea
Using the property of $\vartheta_{\mu\nu}(x)$ defined in Eq.(\ref{Id1}) we get
\bea
\mathcal{Z}_{(1,2)}(\tau,\rho,m,\epsilon)&=&\sum_{\alpha_{1}}\,Q^{2|\alpha_{1}|}\,\frac{\vartheta_{\alpha_{1}\alpha_{1}}(2m)}{\vartheta_{\alpha_{1}\alpha_{1}}(0)}\\\nn
&=&\mathcal{Z}_{(1,1)}(2\tau,\rho,2m,\epsilon)
\eea
\subsection{$(N,M)=(1,M)$ }

The partition function defined in \eqref{Z(N,M)} for $N=1$ has the following expression
\bea
\mathcal{Z}_{(1,M)}(\tau,\rho,\epsilon_{1,2}, m, t)&&=\sum_{\alpha_{1,2,\cdots,M}}\,Q^{|\alpha_{1}|+\cdots+|\alpha_{M}|}\,\prod_{a=1}^{M}\frac{\vartheta_{\alpha_{a}\alpha_{a}}(m)}{\vartheta_{\alpha_{a}\alpha_{a}}(\epsilon_{+})}\nonumber\\
&&\prod_{1\leq a<b\leq M}\frac{\vartheta_{\alpha_{a}\alpha_{b}}(t_{ab}-m)\vartheta_{\alpha_{a}\alpha_{b}}(t_{ab}+m)}{\vartheta_{\alpha_{a}\alpha_{b}}(t_{ab}-\epsilon_{+})\vartheta_{\alpha_{a}\alpha_{b}}(t_{ab}+\epsilon_{+})}
\eea
For $t_{a\,a+1}=m$ we get $t_{ab}=t_{a\,a+1}+t_{a+1\,a+2}+\cdots +t_{a+b-(a+1)\,b}=(b-a)m$. In this case the unrefined $\mathcal{Z}_{(1,M)}$ partition function ($\epsilon_+\rightarrow 0$) becomes
\bea\nonumber
\mathcal{Z}_{(1,M)}(\tau, \rho, t=m,\epsilon)&&=\sum_{\alpha_{1,2,\cdots,M}}Q^{|\alpha_{1}|+\cdots+|\alpha_{M}|}\,\prod_{a=1}^{M}\frac{\vartheta_{\alpha_{a}\alpha_{a}}(m)}{\vartheta_{\alpha_{a}\alpha_{a}}(0)}\nonumber\\&&
\prod_{1\leq a<b\leq M}\frac{\vartheta_{\alpha_{a}\alpha_{b}}((b-a-1)m)\vartheta_{\alpha_{a}\alpha_{b}}((b-a+1)m)}{\vartheta_{\alpha_{a}\alpha_{b}}((b-a)m)^2}
\eea
Since $\vartheta_{\alpha_{a}\alpha_{b}}(0)=0$ for $\alpha_{a}\neq\alpha_{b}$ as shown in the previous section, we get
\bea
\mathcal{Z}_{(1,M)}(\tau, \rho,t=m,\epsilon)&&=\sum_{\alpha_{1}}\,Q^{M|\alpha_{1}|}\,\Big[\frac{\vartheta_{\alpha_{1}\alpha_{1}}(m)}{\vartheta_{\alpha_{1}\alpha_{1}}(0)}\Big]^{M}\prod_{a=1}^{M-1}\prod_{b=a+1}^{M}\frac{\vartheta_{\alpha_{1}\alpha_{1}}((b-a-1)m)\vartheta_{\alpha_{1}\alpha_{1}}((b-a+1)m)}{\vartheta_{\alpha_{1}\alpha_{1}}((b-a)m)^2}\nonumber\\
&&=\sum_{\alpha_{1}}\,Q^{M|\alpha_{1}|}\,\Big[\frac{\vartheta_{\alpha_{1}\alpha_{1}}(m)}{\vartheta_{\alpha_{1}\alpha_{1}}(0)}\Big]^{M}\,\frac{\vartheta_{\alpha_{1}\alpha_{1}}(0)^{M-1}\vartheta_{\alpha_{1}\alpha_{1}}(M\,m)}{\vartheta_{\alpha_{1}\alpha_{1}}(m)^{M}}\nonumber\\\nn
&&=\sum_{\alpha_{1}}Q^{M|\alpha_{1}|}\frac{\vartheta_{\alpha_{1}\alpha_{1}}(M\,m)}{\vartheta_{\alpha_{1}\alpha_{1}}(0)}\nonumber\\
&&=\mathcal{Z}_{(1,1)}(M\,\tau,\rho,Mm,\epsilon)
\eea
This shows self-similarity behaviour of the partition function $\mathcal{Z}_{(1,M)}(\tau, \rho,t=m,\epsilon)$ upto the rescaling of $\tau$ and $m$. In other words as far as the partition function is concerned  the the CY-3fold $X_{N,M}$ is equivalent to the CY-3fold $X_{1,1}$ upto the rescaling of some k\"ahler parameters. This self-similarity structure is actually followed by the partition function $\mathcal{Z}_{(N,M)}(\tau, \rho,t=m,\epsilon)$ for general values of $N$ and $M$ as shown below.
\subsection{General $(N,M)$ }\label{infiniteproductNM}
\begin{figure}[hbt!]
        \center{\includegraphics[width=6in]{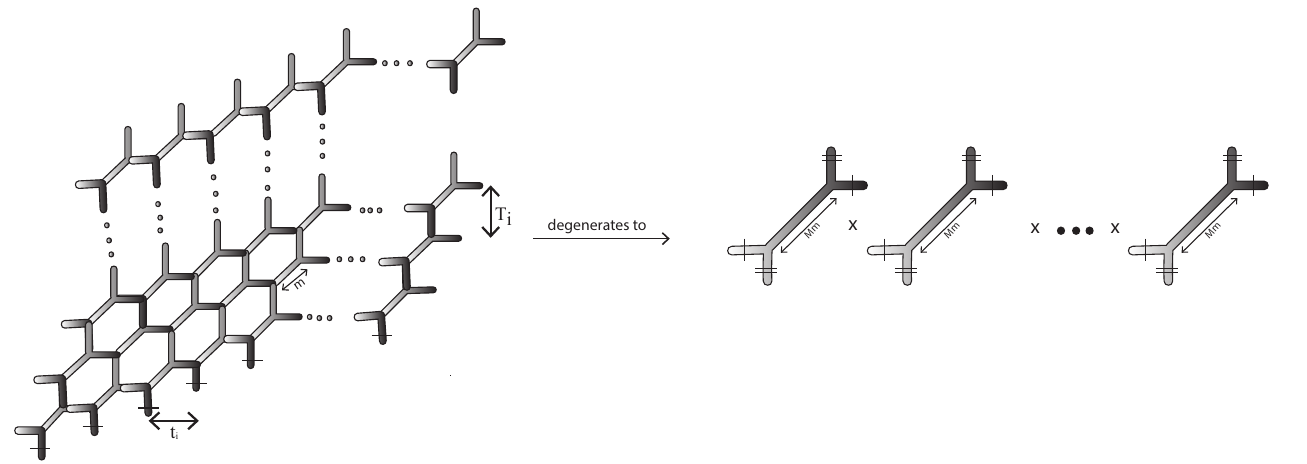}}
        \caption{$\mathcal{Z}_{N,M}$ degenerating to $(\mathcal{Z}_{1,1})^{N}$}
        \label{NM.pdf}
      \end{figure}

%\bea\label{Z(N,M0)}
%\mathcal{Z}_{(N,M)}(\tau, \rho, m, t,\epsilon)&&=\sum_{\alpha_{a}^{i}}\,\prod_{i=1}^{N}Q_{i}^{|\alpha^{(i)}|}\,\prod_{i=1}^{N}\prod_{a=1}^{M}\,\frac{\vartheta_{\alpha_{a}^{i+1}\alpha_{a}^{i}}(m)}{\vartheta_{\alpha_{a}^{i}\alpha_{a}^{i}}(0)}\nonumber\\
%&&\prod_{1\leq a<b\leq M}\prod_{i=1}^{N}\frac{\vartheta_{\alpha_{a}^{i}\alpha_{b}^{i+1}}(t_{ab}-m)\vartheta_{\alpha_{a}^{i+1}\alpha_{b}^{i}}(t_{ab}+m)}{\vartheta_{\alpha_{a}^{i}\alpha_{b}^{i}}(t_{ab})\vartheta_{\alpha_{a}^{i}\alpha_{b}^{i}}(t_{ab})}
%\eea
%As in the previous section we see that
%\bea
%\mathcal{Z}_{(1,M)}(\tau, \rho, m,\epsilon)=\sum_{\alpha_{1}} Q_i^{M|\alpha_{1}|}\,\,\frac{\vartheta_{\alpha_{1}\alpha_{1}}(M\,m)}{\vartheta_{\alpha_{1}\alpha_{1}}(0)}\nonumber
By generalising to the CY-threefold $X_{N,M}$, we get the following result
\bea
\mathcal{Z}_{(N,M)}(\tau, \rho,t=m,\epsilon)&&=\sum_{\alpha^{(i)}_{1}}\prod_{i=1}^N Q_i^{M|\alpha^{(i)}_{1}|}\,\,\frac{\vartheta_{\alpha^{(i)}_{1}\alpha^{(i)}_{1}}(M\,m)}{\vartheta_{\alpha^{(i)}_{1}\alpha^{(i)}_{1}}(0)}\nonumber\\
&&=\sum_{\alpha^{(1)}_{1}} Q_1^{NM|\alpha^{(1)}_{1}|}\,\,\Big(\frac{\vartheta_{\alpha_{1}^{(1)}\alpha^{(1)}_{1}}\,(M\,m)}{\vartheta_{\alpha^{(1)}_{1}\alpha^{(i)}_{1}}(0)}\Big)^N
\eea
So, In general
\bea\label{eq:Deg1}
\mathcal{Z}_{(N,M)}(\tau, \rho, t_{a,a+1}=m,\epsilon)=\mathcal{Z}_{(1,1)}(M\,\tau,\rho,M\,m,\epsilon)^{N}
\eea
This corresponds to degenerating the web diagram of $X_{N,M}$ to the disconnected union of $N$ rescaled web diagrams of $X_{1,1}$ as shown in figure \ref{NM.pdf}.
%\subsection*{Remark 2:}
The CY threefold $X_{1,1}$ has a nice interpretation in terms of the so-called banana curves \cite{bryan2019donaldsonthomas}.
A banana configuration of curves in the CY threefold is a union of three curves $C_i\equiv\mathbb{P}^1$ with the normal bundle given by $\mathcal{O}(-1)\oplus\mathcal{O}(-1)$. Moreover $C_1\cap C_2=C_2\cap C_3=C_3\cap C_1=\{x,y\}$ for distinct point $x,y\in \mbox{CY}$3-fold and there exists a preferred coordinate patch in which $C_i$ are along the coordinate axis. \\
In other words the   topological string partition function $\mathcal{Z}_{X_{N,M}}(\omega,\epsilon)$ is factored  \cite{Haghighat:2018gqf,kawai2000string} into a product of N copies of  $\mathcal{Z}_{X_{1,1}}(\tau,\rho,m)$, where  the later is the topological partition function on a CY threefold with a single banana configuration of curves.
%\begin{figure}[hbt!]
   %     \center{\includegraphics[width=6in]{(N,M).pdf}}
      %  \caption{$\mathcal{Z}_{N,M}$ degenerating to $(\mathcal{Z}_{1,1})^{N}$}
       % \label{NM.pdf}
      %\end{figure}
      
\subsection{Interpreting the factorisation: $\mathcal{Z}_{(M,N)}$  $\to$ $\mathcal{Z}_{(1,1)}^N$ }
Recall that, on a arbitrary point of the K\"ahler cone, the number of independent K\"ahler parameters entering the partition function are
\bea
&\#( T_as)+\#( t_is)+\#(\mbox{intersections})-\#(\mbox{horizontal}\quad \mbox{constraints})-\#(\mbox{vertical}\quad \mbox{constraints})+2\nonumber\\&=(M-1)+(N-1)+MN-(M-1)-(N-1)+2\nonumber\\
&=MN+2
\eea
In general we can have three different series representations \cite{Hohenegger_2017} of $Z_{(M,N)}$ according to whether the toric web diagram of $X_{M,N}$ is sliced into horizontal strips, vertical strips and diagonal strips
\bea
\mathcal{Z}_{(M,N)}(\textbf{t},\textbf{T},\textbf{m},\epsilon_1,\epsilon_2)&=&\mathcal{Z}^{pert}(\textbf{T},\textbf{m})\sum_{\vec{k}}e^{-\vec{k}.\textbf{t}}\mathcal{Z}_{\vec{k}}(\textbf{T},\textbf{m})\nonumber\\
\mathcal{Z}_{(M,N)}(\textbf{t},\textbf{T},\textbf{m},\epsilon_1,\epsilon_2)&=&\mathcal{Z}^{pert}(\textbf{t},\textbf{m})\sum_{\vec{k}}e^{-\vec{k}.\textbf{T}}\mathcal{Z}_{\vec{k}}(\textbf{t},\textbf{m})\nonumber\\
\mathcal{Z}_{(M,N)}(\textbf{t},\textbf{T},\textbf{m},\epsilon_1,\epsilon_2)&=&\mathcal{Z}^{pert}(\textbf{T},\textbf{t})\sum_{\vec{k}}e^{-\vec{k}.\textbf{m}}\mathcal{Z}_{\vec{k}}(\textbf{T},\textbf{t})
\eea
where the K\"ahler parameters $T_i$ from $\bf{T}$$=\{T_1,T_2,...,T_M\}$ represent the distance between vertical lines , $t_i$ from $\bf{t}$$=\{t_1,t_2,...,t_N\}$ represent the distance between horizontal lines and $\bf{m}$ denote the diagonal lines of the web diagram in figure \ref{Fig:WebToric2}.
These expansion have been interpreted as instanton expansions of three gauge theories which are dual to each other.
For these to be consistent expansions it is assumed that there exists a  region of the moduli space of $X_{(M,N)}$ in which either $\textbf{T}$ or $\textbf{t}$ or $\textbf{m}$ become infinite, with all the rest of parameters kept finite. This region of the moduli space corresponds to the weak coupling limit of gauge theories.\\
At the special point in the moduli space where $t_{a,a+1}=m$, we are left with three  independent K\"ahler  parameters $\tau, \rho, m$. Moreover due to the weak coupling expansion $\{\textbf{T}\to \infty\}$, $N$ horizontal strips gets decoupled and we get $Z_{1,1}^N$.
\subsection*{Remark:}
After normalisation by the gauge theory perturbative part, the partition function $Z_{(1,1)}(\tau,\rho,\,m)$ can be written as \cite{Dijkgraaf:2007sw,Bousseau:2020ckw,lockhart2012superconformal}
\bea
\mathcal{Z}_{(1,1)}(\,\tau,\rho,m)&=&e^{\frac{-\pi i (\tau+\rho+m)}{12}}\prod_{(k,l,m)>0}(1-e^{2\pi i(k\tau+l\rho+pm)})^{-c(4kl-p^2)}\nonumber\\
&=&\frac{1}{\Phi_{10}(\tau,\rho,m)^{\frac{1}{24}}}
\eea
where $c(4kl-p^2)$ are the Fourier coefficients of the elliptic genus of $K3$
\bea
\chi(K3,\tau,z)=\sum_{h\ge0,m\in\mathbb{Z}}24c(4h-m^2)e^{2\pi i(h\tau+mz)}
\eea
and $\Phi_{10}(\tau,\rho,m)$ is the unique weight $10$ automorphic form of $Sp(2,\mathbb{Z})$. We have implicit used the fact that the large radius limit (universal part) of the Taub-NUT elliptic genus matches with the elliptic genus of $\mathbb{C}^2$ \cite{Harvey:2014nha}. This allows us to write $Z_{(N,M)}(\tau, \rho, t_{a,a+1}=m)$ in the following way
\bea
\mathcal{Z}_{(N,M)}(\tau, \rho, t_{a,a+1}=m) &=&e^{\frac{-N \pi i (\tau+\rho+m)}{12}}\prod_{(k,l,m)>0}(1-e^{2\pi i(M k\tau+l\rho+pMm)})^{-Nc(4kl-p^2)}\nonumber\\
&=&\frac{1}{\Phi_{10}(M\tau,\rho,Mm)^{\frac{N}{24}}}
\eea

%These three parameters fit nicely into the period matrix of a genus 2 surface.

%\section{Degeneration 2: Splitting Degeneration}\label{splitdegn}
\section{Degeneration 2:Splitting Degeneration}\label{splitdegn}
This degeneration corresponds to turning off the K\"ahler parameters in such a way that the partition function $\mathcal{Z}_{N,M}$ reduces to the partition function  $\mathcal{Z}_{N,M-1}$, upto an overall factor of Dedekind eta function.
 Consider the following partition function
%\bea\label{ZZ(N,M)}
%\mathcal{Z}_{(N,M)}(\tau, \rho, m_{a}\,,\epsilon_{1,2}\,,t_{ab})&&=\sum_{\alpha_{a}^{i}}\,\prod_{i=1}^{N}Q_{i}^{|\alpha^{(i)}|}\,\prod_{i=1}^{N}\prod_{a=1}^{M}\,\frac{\vartheta_{\alpha_{a}^{i+1}\alpha_{a}^{i}}(m_a)}{\vartheta_{\alpha_{a}^{i}\alpha_{a}^{i}}(\epsilon_{+})}\nonumber\\
%&&\prod_{1\leq a<b\leq M}\prod_{i=1}^{N}\frac{\vartheta_{\alpha_{a}^{i}\alpha_{b}^{i+1}}(t_{ab}-m_a)\vartheta_{\alpha_{a}^{i+1}\alpha_{b}^{i}}(t_{ab}+m_b)}{\vartheta_{\alpha_{a}^{i}\alpha_{b}^{i}}(t_{ab}-\epsilon_{+})\vartheta_{\alpha_{a}^{i}\alpha_{b}^{i}}(t_{ab}+\epsilon_{+})}
%\eea
% For $N=1$ the above defined partition function reduces to
%\bea
%\mathcal{Z}_{(1,M)}(\tau, \rho, m_{a}\,,\epsilon_{1,2}\,,t_{ab})&&=\sum_{\alpha_{1,2,\cdots,M}}\,Q^{|\alpha_{1}|+\cdots+|\alpha_{M}|}\,\prod_{a=1}^{M}\frac{\vartheta_{\alpha_{a}\alpha_{a}}(m_a)}{\vartheta_{\alpha_{a}\alpha_{a}}(\epsilon_{+})}\nonumber\\
%&&\prod_{1\leq a<b\leq M}\frac{\vartheta_{\alpha_{a}\alpha_{b}}(t_{ab}-m_a)\vartheta_{\alpha_{a}\alpha_{b}}(t_{ab}+m_b)}{\vartheta_{\alpha_{a}\alpha_{b}}(t_{ab}-\epsilon_{+})\vartheta_{\alpha_{a}\alpha_{b}}(t_{ab}+\epsilon_{+})}\nonumber
%\eea
\bea\label{ZZ(N,M)}
\mathcal{Z}_{(N,M)}(\tau, \rho, m_{a}\,,\epsilon_{1,2},\widetilde{t}_{ab})&&=\sum_{\alpha_{a}^{i}}\,\prod_{i=1}^{N}Q_{i}^{|\alpha^{(i)}|}\,\prod_{i=1}^{N}\prod_{a=1}^{M}\,\frac{\vartheta_{\alpha_{a}^{i+1}\alpha_{a}^{i}}(m_a)}{\vartheta_{\alpha_{a}^{i}\alpha_{a}^{i}}(\epsilon_{+})}\nonumber\\
&&\prod_{1\leq a<b\leq M}\prod_{i=1}^{N}\frac{\vartheta_{\alpha_{a}^{i}\alpha_{b}^{i+1}}(\widetilde{t}_{ab})\vartheta_{\alpha_{a}^{i+1}\alpha_{b}^{i}}(\widetilde{t}_{ab}+m_a+m_b)}{\vartheta_{\alpha_{a}^{i}\alpha_{b}^{i}}(\widetilde{t}_{ab}+m_a-\epsilon_{+})\vartheta_{\alpha_{a}^{i}\alpha_{b}^{i}}(\widetilde{t}_{ab}+m_b+\epsilon_{+})}.
\eea
In the above partition function \eqref{ZZ(N,M)}
$$\widetilde{t}_{ab}=\widetilde{t}_{a\,a+1}+m_{a+1}+\widetilde{t}_{a+1\,a+2}+\cdots+m_{b-1}+\widetilde{t}_{b-1\,b}.$$ For $N=1$ the above defined partition function reduces to
 \bea
\mathcal{Z}_{(1,M)}(\tau, \rho, m_{a}\,,\epsilon_{1,2}\,,\widetilde{t_{ab}})&&=\sum_{\alpha_{1,2,\cdots,M}}\,Q^{|\alpha_{1}|+\cdots+|\alpha_{M}|}\,\prod_{a=1}^{M}\frac{\vartheta_{\alpha_{a}\alpha_{a}}(m_a)}{\vartheta_{\alpha_{a}\alpha_{a}}(\epsilon_{+})}\nonumber\\
&&\prod_{1\leq a<b\leq M}\frac{\vartheta_{\alpha_{a}\alpha_{b}}(\widetilde{t_{ab}})\vartheta_{\alpha_{a}\alpha_{b}}(\widetilde{t_{ab}}+m_a+m_b)}{\vartheta_{\alpha_{a}\alpha_{b}}(\widetilde{t_{ab}}+m_a-\epsilon_{+})\vartheta_{\alpha_{a}\alpha_{b}}(\widetilde{t_{ab}}+m_b+\epsilon_{+})}\nonumber
\eea
\subsection*{Remark:}
Note that $\sum_{ \mu}Q^{|\mu|}=\frac{e^{\frac{\pi i }{12}}}{\eta(\tau)}$. This factor appears in the degeneration limit as discussed below.
\subsection{$(N,M)=(1,2)$ }
Let us consider the partition function for $N=1$ and $M=2$ in the unrefined case ($\epsilon_1=-\epsilon_2=\epsilon$),
\bea\label{sp12}
\mathcal{Z}_{(1,2)}(\tau,\rho,m_{1,2},\widetilde{t_{12}},\epsilon)=\sum_{\alpha_{1,2}}\,Q^{|\alpha_{1}|+|\alpha_{2}|}\,\frac{\vartheta_{\alpha_{1}\alpha_{1}}(m_1)\vartheta_{\alpha_{2}\alpha_{2}}(m_2)}{\vartheta_{\alpha_{1}\alpha_{1}}(0)\vartheta_{\alpha_{2}\alpha_{2}}(0)} \frac{\vartheta_{\alpha_{1}\alpha_{2}}(\widetilde{t_{12}})\vartheta_{\alpha_{1}\alpha_{2}}(\widetilde{t_{12}}+m_1+m_2)}{\vartheta_{\alpha_{1}\alpha_{2}}(\widetilde{t_{12}}+m_1)\vartheta_{\alpha_{1}\alpha_{2}}(\widetilde{t_{12}}+m_2)}
\eea
$\bullet \quad \mathbf{m_1\rightarrow 0}$ or $\mathbf{m_2\rightarrow 0}$:\\
When we take $m_1=0$ in the partition function \eqref{sp12}, the terms in the numerator and denominator becomes same, therefore they cancel out each other. Then \eqref{sp12} reduces to the multiple of $Z_{(1,1)}$ as:
\bea
\mathcal{Z}_{(1,2)}(\tau,\rho, m_2, \widetilde{t_{12}},\epsilon)&=\sum\limits_{\alpha_{1,2}}\,Q^{|\alpha_{1}|+|\alpha_{2}|}\,\frac{\vartheta_{\alpha_{2}\alpha_{2}}(m_2)}{\vartheta_{\alpha_{2}\alpha_{2}}(0)}\nonumber\\
&=\sum\limits_{\alpha_1}Q^{|\alpha_1|}\mathcal{Z}_{(1,1)}(\tau,\rho,m_2,\epsilon)\nonumber
\eea
%\begin{figure}[hbt!]
%        \center{\includegraphics[width=6in]{sap12.pdf}}
%    %    \caption{\label{fig:X12} tessellation of Newton polygons and web diagram of $X_{1,2}$}
%      \end{figure}
Same result follows for the case when we take $m_2=0$ in \eqref{sp12} i.e,
\bea
\mathcal{Z}_{(1,2)}(\tau, \rho, m_1, \widetilde{t_{12}}, \epsilon)=\sum_{\alpha_2}Q^{|\alpha_2|}\mathcal{Z}_{(1,1)}(\tau, \rho, m_2, \epsilon)\nonumber
\eea
$\bullet$ \quad$\widetilde{\mathbf{t}_{12}}\rightarrow 0:$\\
In the limit $\widetilde{t_{12}}\rightarrow 0$, \eqref{sp12} is:
\bea
\mathcal{Z}_{(1,2)}(\tau,\rho,m_{1,2},\epsilon)=\sum_{\alpha_{1,2}}\,Q^{|\alpha_{1}|+|\alpha_{2}|}\,\frac{\vartheta_{\alpha_{1}\alpha_{1}}(m_1)\vartheta_{\alpha_{2}\alpha_{2}}(m_2)}{\vartheta_{\alpha_{1}\alpha_{1}}(0)\vartheta_{\alpha_{2}\alpha_{2}}(0)} \frac{\vartheta_{\alpha_{1}\alpha_{2}}(0)\vartheta_{\alpha_{1}\alpha_{2}}(m_1+m_2)}{\vartheta_{\alpha_{1}\alpha_{2}}(m_1)\vartheta_{\alpha_{1}\alpha_{2}}(m_2)}
\eea
Again, the presence of $\vartheta_{\alpha_{1}\alpha_{1}}(0)$ force contribution only from the same partition and we get the following:
$$\mathcal{Z}_{(1,2)}(\tau,\rho,m_{1,2},\epsilon)=\mathcal{Z}_{(1, 1)}(2\tau, \rho, m_{1}+m_2, \epsilon)$$

\begin{figure}[hbt!]
        \center{\includegraphics[width=6in]{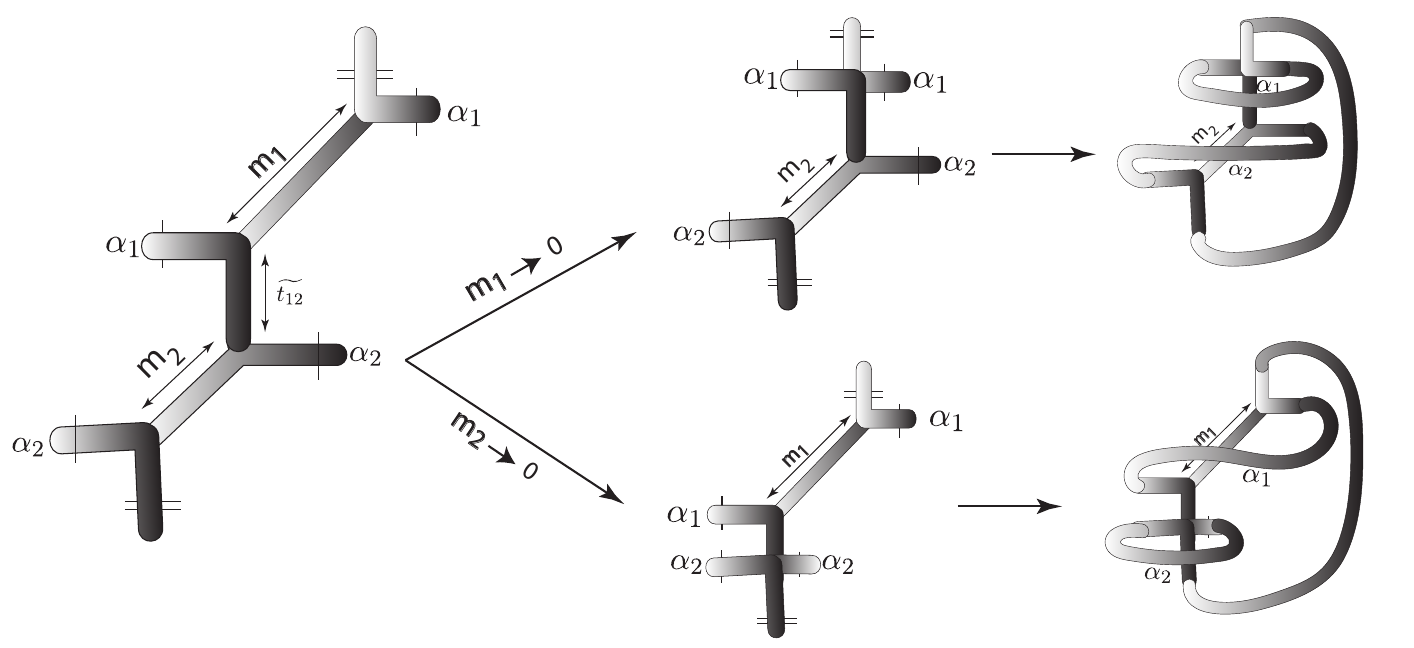}}
    %    \caption{\label{fig:X12} tessellation of Newton polygons and web diagram of $X_{1,2}$}
      \caption{Two possible degenerations of the partition function $Z_{1,2}$. The third column depicts  3D/non-planar structure of the mirror curves}
\label{Fig:WebToricZ12}
      \end{figure}

\subsection{$(N,M)=(1,3)$ }
%For $M=3$ we have similar degeneration
% \bea\nonumber
%&&\mathcal{Z}_{(1,3)}(\tau, \rho, m_{1},m_2,m_3, t_{12},t_{23},t_{13},\epsilon)=\sum_{\alpha_{1,2,3}}\,Q^{|\alpha_{1}|+|\alpha_{2}|+|\alpha_{3}|}\,\frac{\vartheta_{\alpha_{1}\alpha_{1}}(m_1)\vartheta_{\alpha_{2}\alpha_{2}}(m_2)\vartheta_{\alpha_{3}\alpha_{3}}(m_3)}{\vartheta_{\alpha_{1}\alpha_{1}}(0)\vartheta_{\alpha_{2}\alpha_{2}}(0)\vartheta_{\alpha_{3}\alpha_{3}}(0)} \\&&\frac{\vartheta_{\alpha_{1}\alpha_{2}}(t_{12}-m_1)\vartheta_{\alpha_{1}\alpha_{2}}(t_{12}+m_2)}{\vartheta_{\alpha_{1}\alpha_{2}}(t_{12})\vartheta_{\alpha_{1}\alpha_{2}}(t_{12})}
%\frac{\vartheta_{\alpha_{2}\alpha_{3}}(t_{23}-m_2)\vartheta_{\alpha_{2}\alpha_{3}}(t_{23}+m_3)}{\vartheta_{\alpha_{2}\alpha_{3}}(t_{23})\vartheta_{\alpha_{2}\alpha_{3}}(t_{23})}
%\frac{\vartheta_{\alpha_{1}\alpha_{3}}(t_{13}-m_1)\vartheta_{\alpha_{1}\alpha_{3}}(t_{13}+m_3)}{\vartheta_{\alpha_{1}\alpha_{3}}(t_{13})\vartheta_{\alpha_{1}\alpha_{3}}(t_{13})}\nonumber\\
%\eea
%
Similarly consider the partition function $\mathcal{Z}_{(1,3)}(\tau, \rho, m_{1,2,3}, \widetilde{t}_{a,b},\epsilon)$ 
\bea\label{Z13}
&&\mathcal{Z}_{(1,3)}(\tau, \rho, m_{1,2,3}, \widetilde{t}_{a,b},\epsilon)=\sum\limits_{\alpha_{1,2,3}}\,Q^{\sum\limits_{k=1}^3|\alpha_{k}|}\,\left(\prod_{k=1}^3\frac{\vartheta_{\alpha_{k}\alpha_{k}}(m_k)}{\vartheta_{\alpha_{k}\alpha_{k}}(0)} \right) \frac{\vartheta_{\alpha_{1}\alpha_{2}}(\widetilde{t_{12}})\vartheta_{\alpha_{1}\alpha_{2}}(\widetilde{t_{12}}+m_1+m_2)}{\vartheta_{\alpha_{1}\alpha_{2}}(m_1+\widetilde{t_{12}})\vartheta_{\alpha_{1}\alpha_{2}}(m_2+\widetilde{t_{12}})} \nonumber\\&&\quad\quad\quad\quad\frac{\vartheta_{\alpha_{2}\alpha_{3}}(\widetilde{t_{23}})\vartheta_{\alpha_{2}\alpha_{3}}(\widetilde{t_{23}}+m_2+m_3)}{\vartheta_{\alpha_{2}\alpha_{3}}(\widetilde{t_{23}}+m_2)\vartheta_{\alpha_{2}\alpha_{3}}(\widetilde{t_{23}}+m_3)}
\frac{\vartheta_{\alpha_{1}\alpha_{3}}(\widetilde{t_{13}})\vartheta_{\alpha_{1}\alpha_{3}}(\widetilde{t_{13}}+m_1+m_3)}{\vartheta_{\alpha_{1}\alpha_{3}}(\widetilde{t_{13}}+m_1)\vartheta_{\alpha_{1}\alpha_{3}}(\widetilde{t_{13}}+m_3)}
\eea
\begin{figure}[h!]
        \center{\includegraphics[width=2in]{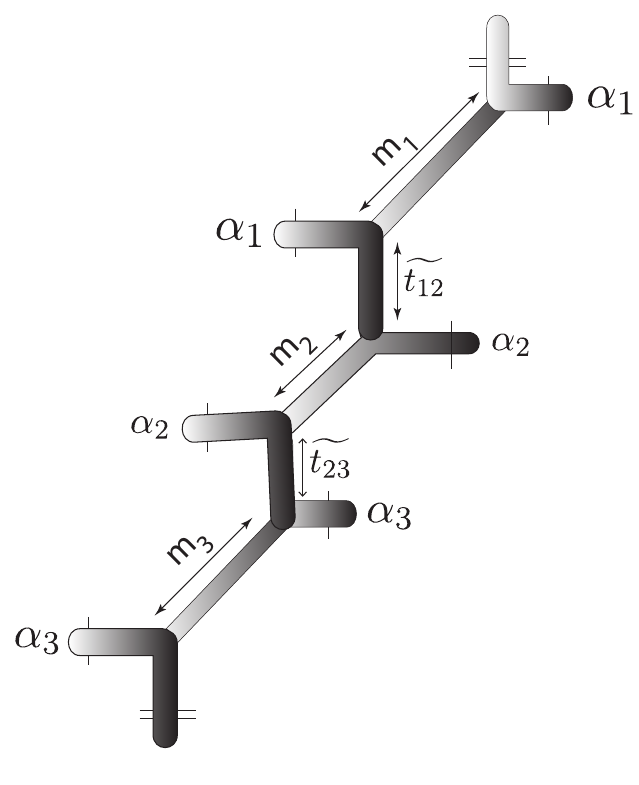}}
    %    \caption{\label{fig:X12} tessellation of Newton polygons and web diagram of $X_{1,2}$}
    \caption{$Z_{13}$}
\label{Fig:WebToricZ13}
      \end{figure}

\begin{figure}[h!]
        \center{\includegraphics[width=6in]{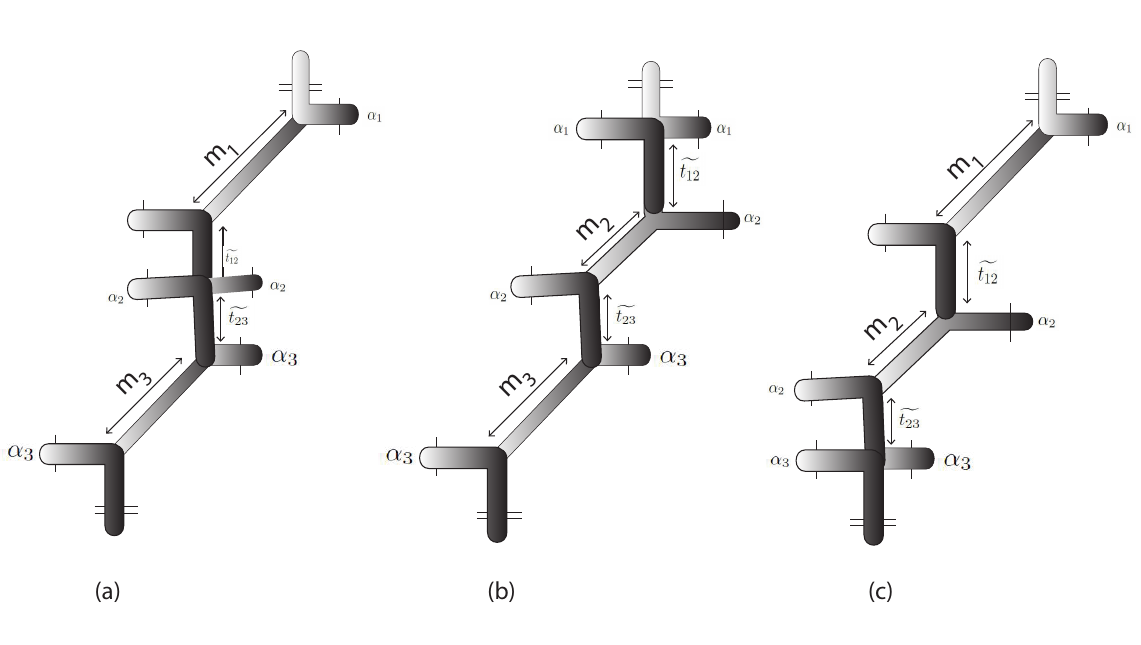}}
    %    \caption{\label{fig:X12} tessellation of Newton polygons and web diagram of $X_{1,2}$}
    \caption{Three possible degenerations of the partition function $Z_{1,3}$. }
\label{Fig:DegWebToricZ13}
      \end{figure}
%Here, all $m_i'$s $i=1,2,3$ are different, $t_{13}= t_{12}+ t_{23}$ with all $t_{ab}$'s $a=1,2$ and $b=2, 3$ different. To make the calculation simple we rewrite the partition function in the following way after making the redefinitions $\widetilde{t}_{12}=t_{12}-m_1$,$\widetilde{t}_{23}=t_{23}-m_2$ and $\widetilde{t}_{13}=t_{13}-m_1$:
\newpage
Remember here all $m_i'$s $i=1,2,3$ are different, and $\widetilde{t_{13}}= \widetilde{t_{12}}+ m_2+ \widetilde{t_{23}}.$\\\\
$\bullet$\quad $\mathbf{m_3\mapsto0}:$\\
When $m_3$ approaches to zero in \eqref{Z13} it takes the following form:
\bea
\mathcal{Z}_{(1,3)}(\tau, \rho, m_{1,2}, \widetilde{t_{12}},\epsilon)&&=\sum_{\alpha_{1,2,3}}\,Q^{|\alpha_{1}|+|\alpha_{2}|+|\alpha_{3}|}\,\frac{\vartheta_{\alpha_{1}\alpha_{1}}(m_1)\vartheta_{\alpha_{2}\alpha_{2}}(m_2)}{\vartheta_{\alpha_{1}\alpha_{1}}(0)\vartheta_{\alpha_{2}\alpha_{2}}(0)} \frac{\vartheta_{\alpha_{1}\alpha_{2}}(\widetilde{t_{12}})\vartheta_{\alpha_{1}\alpha_{2}}(\widetilde{t_{12}}+m_1+m_2)}{\vartheta_{\alpha_{1}\alpha_{2}}(m_1+\widetilde{t_{12}})\vartheta_{\alpha_{1}\alpha_{2}}(m_2+\widetilde{t_{12}})}\nonumber\\
&&=\sum_{\alpha_3}Q^{|\alpha_{3}|}\, \mathcal{Z}_{(1,2)}(\tau, \rho, m_{1,2}, \widetilde{t_{12}},\epsilon)\nonumber
\eea
Thus $Z_{(1,3)}\rightarrow Z_{(1,2)}$.\\\newpage
$\bullet$\quad $\mathbf{m_2\mapsto0}:$\\
Similarly
\bea\nonumber
\mathcal{Z}_{(1,3)}(\tau, \rho, m_{1,2,3}, \widetilde{t_{13}},\epsilon)\quad\underrightarrow{m_2=0}\quad\sum_{\alpha_2}Q^{|\alpha_{2}|}\, \mathcal{Z}_{(1,2)}(\tau, \rho, m_{1,3}, \widetilde{t_{13}},\epsilon)
\eea
%$\mathbf{m_1=0}:$
and \\
$\bullet$\quad $\mathbf{m_1\mapsto0}:$\\
\bea\nonumber
\mathcal{Z}_{(1,3)}(\tau, \rho, m_{1,2,3}, \widetilde{t_{23}},\epsilon)\quad\underrightarrow{m_1=0}\quad\sum_{\alpha_1}Q^{|\alpha_{1}|}\,\mathcal{ Z}_{(1,2)}(\tau, \rho, m_{2,3}, \widetilde{t_{23}},\epsilon)
\eea
Hence in all these three cases when any $m_1, m_2$ or $m_3$ is zero $\mathcal{Z}_{(1,3)}$ reduces to the case of $\mathcal{Z}_{(1,2)}$ upto some factor.
Moreover same degeneration of $\mathcal{Z}_{1,3}$ results if one takes the limit   $\widetilde{t_{ab}}\rightarrow 0$ for any $a,b$ i.e., $\mathcal{Z}_{(1,3)}\to \mathcal{Z}_{(1,2)}$.\\
$\bullet$\quad $\mathbf{\widetilde{t}_{ab}\mapsto0}:$\\
\bea\nonumber
\mathcal{Z}_{(1,3)}(\tau, \rho, m_{1,2,3}, \widetilde{t_{23}},\epsilon)\quad\underrightarrow{\widetilde{t}_{ab}=0}\quad\sum_{\alpha}Q^{|\alpha|}\,\mathcal{ Z}_{(1,2)}(\tau, \rho, m_{1,2,3},\epsilon)
\eea
\subsection{$(N,M)=(2,3)$ }
Previous subsections discuss the cases when $N=1$ and now we generalize to the case of $N=2$. Explicitly the partition function is of the form
\bea
\mathcal{Z}_{(2,3)}(\tau, \rho, m_{1,2,3},\epsilon_{1,2},\widetilde{t}_{ab})&=&\sum_{\alpha_{a}^{i}}\,\prod_{i=1}^{2}Q_{i}^{|\alpha^{(i)}|}\,\prod_{i=1}^{2}\prod_{a=1}^{3}\,\frac{\vartheta_{\alpha_{a}^{i+1}\alpha_{a}^{i}}(m_a)}{\vartheta_{\alpha_{a}^{i}\alpha_{a}^{i}}(0)}\nonumber\\
&\times&\prod_{1\leq a<b\leq 3}\prod_{i=1}^{2}\frac{\vartheta_{\alpha_{a}^{i}\alpha_{b}^{i+1}}(\widetilde{t_{ab}})\vartheta_{\alpha_{a}^{i+1}\alpha_{b}^{i}}(\widetilde{t_{ab}}+m_a+m_b)}{\vartheta_{\alpha_{a}^{i}\alpha_{b}^{i}}(\widetilde{t_{ab}}+m_a-\epsilon_{+})\vartheta_{\alpha_{a}^{i}\alpha_{b}^{i}}(\widetilde{t_{ab}}+m_b+\epsilon_{+})}
\eea
For the unrefined case $\epsilon_1=-\epsilon_2=\epsilon$, we consider the degenerate limit  $m_3=0$.  Using the identity (\ref{Id1})  we get
\bea
&&\mathcal{Z}_{(2,3)}(\tau, \rho, m_{1,2},m_3=0,\widetilde{t}_{ab},\epsilon)=\sum_{\alpha^{1},\alpha^{2}, \alpha_3^{(1)}}\,Q_{1}^{|\alpha^{(1)}|}Q_{2}^{|\alpha^{(2)}|}(Q_{1}Q_{2})^{|\alpha_3^{(1)}|}\,\frac{\vartheta_{\alpha_{1}^{(2)}\alpha_1^{(1)}}(m_1)\vartheta_{\alpha_{2}^{(2)}\alpha_2^{(1)}}(m_2)}{\vartheta_{\alpha_{1}^{(1)}\alpha_1^{(1)}}(0)\vartheta_{\alpha_{2}^{(1)}\alpha_2^{(1)}}(0)}
\nonumber\\&&
\frac{\vartheta_{\alpha_{1}^{(1)}\alpha_1^{(2)}}(m_1)\vartheta_{\alpha_{2}^{(1)}\alpha_2^{(2)}}(m_2)}{\vartheta_{\alpha_{1}^{(2)}\alpha_1^{(2)}}(0)\vartheta_{\alpha_{2}^{(2)}\alpha_2^{(2)}}(0)} \frac{\vartheta_{\alpha_{1}^{(1)}\alpha_2^{(2)}}(\widetilde{t_{12}})\vartheta_{\alpha_{1}^{(2)}\alpha_2^{(1)}}(\widetilde{t_{12}}+m_1+m_2)}{\vartheta_{\alpha_{1}^{(1)}\alpha_2^{(1)}}(\widetilde{t_{12}}+m_1)\vartheta_{\alpha_{1}^{(1)}\alpha_2^{(1)}}(\widetilde{t_{12}}+m_2)}
\frac{\vartheta_{\alpha_{1}^{(2)}\alpha_2^{(1)}}(\widetilde{t_{12}})\vartheta_{\alpha_{1}^{(1)}\alpha_2^{(2)}}(\widetilde{t_{12}}+m_1+m_2)}{\vartheta_{\alpha_{1}^{(2)}\alpha_2^{(2)}}(\widetilde{t_{12}}+m_1)\vartheta_{\alpha_{1}^{(2)}\alpha_2^{(2)}}(\widetilde{t_{12}}+m_2)}\nonumber
\eea
%\bea
%&Z_{(2,3)}(\tau, \rho, m_{2,3}, t_{ab},\epsilon)=\sum\limits_{\alpha_{1}^1,\alpha_1^2}Q^{\alpha_{1}^1+\alpha_1^2}\bigg(\frac{\vartheta_{\alpha_{1}^1\alpha_1^2}(0)}{\vartheta_{\alpha_{1}^1\alpha_1^1}(0)}\bigg)^2\bigg[\sum\limits_{\alpha_{2}^1,\alpha_2^2,\alpha_{3}^1,\alpha_3^2}Q^{\alpha_{2}^1+\alpha_2^2+\alpha_{3}^1+\alpha_3^2}\prod\limits_{i=1}^{2}\prod\limits_{a=2}^{3}\,\frac{\vartheta_{\alpha_{a}^{i+1}\alpha_{a}^{i}}(m_a)}{\vartheta_{\alpha_{a}^{i}\alpha_{a}^{i}}(0)}\nonumber\\&
%\prod\limits_{2\leq a<b\leq 3}\prod\limits_{i=1}^{2}\frac{\vartheta_{\alpha_{2}^{i}\alpha_{3}^{i+1}}(t_{23}-m_2)\vartheta_{\alpha_{2}^{i+1}\alpha_{3}^{i}}(t_{23}+m_3)}{\vartheta_{\alpha_{2}^{i}\alpha_{3}^{i}}(t_{23})\vartheta_{\alpha_{2}^{i}\alpha_{3}^{i}}(t_{23})}\bigg]
%\bigg(\frac{\vartheta_{\alpha_{1}^{2}\alpha_{2}^{1}}(t_{12}+m_2)\vartheta_{\alpha_{1}^{1}\alpha_{2}^{2}}(t_{12}+m_2)}{\vartheta_{\alpha_{1}^{1}\alpha_{2}^{1}}(t_{12})\vartheta_{\alpha_{1}^{2}\alpha_{2}^{2}}(t_{12})}
%\frac{\vartheta_{\alpha_{1}^{2}\alpha_{3}^{1}}(t_{13}+m_3)\vartheta_{\alpha_{1}^{1}\alpha_{3}^{2}}(t_{13}+m_3)}{\vartheta_{\alpha_{1}^{1}\alpha_{3}^{1}}(t_{13})\vartheta_{\alpha_{1}^{2}\alpha_{3}^{2}}(t_{13})}\bigg)\nonumber
%\eea
Recognizing the  $\mathcal{Z}_{(2,2)}(\tau, \rho, m_{1,2}, t_{ab},\epsilon)$ part, the last expression can be written more succinctly as
\bea
\mathcal{Z}_{(2,3)}(\tau, \rho, m_{1,2},m_3=0, \widetilde{t}_{ab},\epsilon)&=&\sum_{\alpha_{3}^{(1)}}Q_1^{\alpha_{3}^{(1)}}Q_2^{\alpha_{3}^{(1)}}\mathcal{Z}_{(2,2)}(\tau, \rho, m_{1,2}, \widetilde{t}_{ab},\epsilon)
\eea
%where  $Z^{\alpha_{2}^1,\alpha_2^2,\alpha_{3}^1,\alpha_3^2}_{(2,2)}(\tau, \rho, m_{2,3}, t_{23},\epsilon)$ is defined by
%\bea\label{eq:Nge2M}
%Z_{(2,2)}(\tau, \rho, m_{2,3}, t_{23},\epsilon):=\sum_{\alpha_{2}^1,\alpha_2^2,\alpha_{3}^1,\alpha_3^2}Z^{\alpha_{2}^1,\alpha_2^2,\alpha_{3}^1,\alpha_3^2}_{(2,2)}(\tau, \rho, m_{2,3}, t_{23},\epsilon)
%\eea
Similar degenerations follow by taking the limit  $m_2=0$ or $m_1=0$.

\subsection{General $(N,M)$  }
%The expression for $\mathcal{Z}_{N,M}$ can be simplified, for the unrefined  case $\epsilon_1=-\epsilon_2=\epsilon$, as follows
%\bea\label{ZZ(N,M)2}
%\mathcal{Z}_{(N,M)}(\tau, \rho, m_{q}\,,\epsilon\,,t_{ab})&=&\sum_{\alpha_{a}^{i}}\,\prod_{i=1}^{N}Q_{i}^{|\alpha^{(i)}|}\,\prod_{i=1}^{N}\prod_{a=1}^{M}\,\frac{\vartheta_{\alpha_{a}^{i+1}\alpha_{a}^{i}}(m_a)}{\vartheta_{\alpha_{a}^{i}\alpha_{a}^{i}}(0)}\nonumber\\
%&\times&\prod_{1\leq a<b\leq M}\prod_{i=1}^{N}\frac{\vartheta_{\alpha_{a}^{i}\alpha_{b}^{i+1}}(t_{ab}-m_a)\vartheta_{\alpha_{a}^{i+1}\alpha_{b}^{i}}(t_{ab}+m_b)}{\vartheta_{\alpha_{a}^{i}\alpha_{b}^{i}}(t_{ab})\vartheta_{\alpha_{a}^{i}\alpha_{b}^{i}}(t_{ab})}\nonumber\\
%&=& \sum_{\alpha_{a}^{i}}\,\prod_{i=1}^{N}Q_{i}^{|\alpha^{(i)}|}\,\prod_{i=1}^{N}\prod_{a=1}^{M}\,\frac{\vartheta_{\alpha_{a}^{i+1}\alpha_{a}^{i}}(m_a)}{\vartheta_{\alpha_{a}^{i}\alpha_{a}^{i}}(0)}\nonumber\\
%&\times&
%\prod_{a=1}^{M-1}\prod_{b=a+1}^M\frac{\theta_{\alpha_a^i\alpha_b^{i+1}}(\widetilde{t}_{aa+1}+\widetilde{t}_{a+1a+2}+...+\widetilde{t}_{b-1b}+m_{a+1}+...+m_{b-1})}{\theta_{\alpha_a^i\alpha_b^{i}}(\widetilde{t}_{aa+1}+\widetilde{t}_{a+1a+2}+...+\widetilde{t}_{b-1b}+m_a+m_{a+1}+...+m_{b-1})}\nonumber\\
%&\times&\frac{\theta_{\alpha_a^i\alpha_b^{i+1}}(\widetilde{t}_{aa+1}+\widetilde{t}_{a+1a+2}+...+\widetilde{t}_{b-1b}+m_a+m_{a+1}+...+m_{b-1}+m_b)}{\theta_{\alpha_a^i\alpha_b^{i}}(\widetilde{t}_{aa+1}+\widetilde{t}_{a+1a+2}+...+\widetilde{t}_{b-1b}+m_{a+1}+...+m_{b-1}+m_b)}\nonumber\\
%\eea
The previous sections discuss the cases when $N$ was taken equal to one. In this section we generalize the argument to generic values of $M$ and $N.$
For the unrefined  case $\epsilon_1=-\epsilon_2=\epsilon$
\bea\label{ZZ(N,M)2}
\mathcal{Z}_{(N,M)}(\tau, \rho, m_{a},\epsilon_{1,2},\widetilde{t}_{ab})&=&\sum_{\alpha_{a}^{i}}\,\prod_{i=1}^{N}Q_{i}^{|\alpha^{(i)}|}\,\prod_{i=1}^{N}\prod_{a=1}^{M}\,\frac{\vartheta_{\alpha_{a}^{i+1}\alpha_{a}^{i}}(m_a)}{\vartheta_{\alpha_{a}^{i}\alpha_{a}^{i}}(0)}\nonumber\\
&\times&\prod_{1\leq a<b\leq M}\prod_{i=1}^{N}\frac{\vartheta_{\alpha_{a}^{i}\alpha_{b}^{i+1}}(\widetilde{t_{ab}})\vartheta_{\alpha_{a}^{i+1}\alpha_{b}^{i}}(\widetilde{t_{ab}}+m_a+m_b)}{\vartheta_{\alpha_{a}^{i}\alpha_{b}^{i}}(\widetilde{t_{ab}}+m_a)\vartheta_{\alpha_{a}^{i}\alpha_{b}^{i}}(\widetilde{t_{ab}}+m_b)}
\eea

%\bea\label{ZZ(N,M)2}
%\mathcal{Z}_{(N,M)}(\tau, \rho, m_{i}\,\widetilde{t}_{ab},\epsilon)&=&\sum_{\alpha_{a}^{i}}\,\prod_{i=1}^{N}Q_{i}^{|\alpha^{(i)}|}\,\prod_{i=1}^{N}\prod_{a=1}^{M}\,\frac{\vartheta_{\alpha_{a}^{i+1}\alpha_{a}^{i}}(m_a)}{\vartheta_{\alpha_{a}^{i}\alpha_{a}^{i}}(0)}\nonumber\\
%&\times&\prod_{1\leq a<b\leq M}\prod_{i=1}^{N}\frac{\vartheta_{\alpha_{a}^{i}\alpha_{b}^{i+1}}(\widetilde{t_{ab}})\vartheta_{\alpha_{a}^{i+1}\alpha_{b}^{i}}(\widetilde{t_{ab}}+m_a+m_b)}{\vartheta_{\alpha_{a}^{i}\alpha_{b}^{i}}(\widetilde{t}_{ab})\vartheta_{\alpha_{a}^{i}\alpha_{b}^{i}}(\widetilde{t}_{ab})}\nonumber\\
%\eea
%Since, $\widetilde{t}_{ab}=\widetilde{t}_{a\,a+1}+m_{a+1}+\widetilde{t}_{a+1\,a+2}+\cdots+m_{b-1}+\widetilde{t}_{b-1\,b}$.
\bea
\mathcal{Z}_{(N,M)}(\tau, \rho, m_{a},\widetilde{t}_{ab},\epsilon)&=& \sum_{\alpha_{a}^{i}}\,\prod_{i=1}^{N}Q_{i}^{|\alpha^{(i)}|}\,\prod_{i=1}^{N}\prod_{a=1}^{M}\,\frac{\vartheta_{\alpha_{a}^{i+1}\alpha_{a}^{i}}(m_a)}{\vartheta_{\alpha_{a}^{i}\alpha_{a}^{i}}(0)}\nonumber\\
&\times&
\prod_{a=1}^{M-1}\prod_{b=a+1}^M\frac{\vartheta_{\alpha_a^i\alpha_b^{i+1}}(\widetilde{t}_{a\,a+1}+m_{a+1}+\widetilde{t}_{a+1\,a+2}+\cdots+m_{b-1}+\widetilde{t}_{b-1\,b})}{\vartheta_{\alpha_a^i\alpha_b^{i}}(\widetilde{t}_{a\,a+1}+m_{a+1}+\widetilde{t}_{a+1\,a+2}+\cdots+m_{b-1}+\widetilde{t}_{b-1\,b})}\nonumber\\
&\times&\frac{\vartheta_{\alpha_a^i\alpha_b^{i+1}}(\widetilde{t}_{a\,a+1}+m_{a+1}+\widetilde{t}_{a+1\,a+2}+\cdots+m_{b-1}+\widetilde{t}_{b-1\,b}+m_a+m_b)}{\vartheta_{\alpha_a^i\alpha_b^{i}}(\widetilde{t}_{a\,a+1}+m_{a+1}+\widetilde{t}_{a+1\,a+2}+\cdots+m_{b-1}+\widetilde{t}_{b-1\,b})}\nonumber\\
\eea
Specializing to  $N=1$, $Q_i=Q$ and in the limit $m_1=0$ the last expression reduces to
\bea\label{concatenatedM5}
\mathcal{Z}_{(1,M)}(\tau, \rho, m_{a},\widetilde{t}_{ab},\epsilon)
%&=& \sum_{\alpha_{}}\,Q^{|\alpha|} \sum_{\alpha_{a}}\,Q^{|\alpha_a|}\,\prod_{a=2}^{M}\,\frac{\vartheta_{\alpha_{a}^{i+1}\alpha_{a}^{i}}(m_a)}{\vartheta_{\alpha_{a}^{i}\alpha_{a}^{i}}(0)}\nonumber\\
%&\times&
%\prod_{a=2}^{M-1}\prod_{b=a+1}^M\nonumber\\
%&\times&\bigg(\frac{\theta_{\alpha_a\alpha_b}(\widetilde{t}_{aa+1}+\widetilde{t}_{a+1a+2}+...+\widetilde{t}_{b-1b}+m_{a+1}+...+m_{b-1})}{\theta_{\alpha_a\alpha_b}(\widetilde{t}_{aa+1}+\widetilde{t}_{a+1a+2}+...+\widetilde{t}_{b-1b}+m_a+m_{a+1}+...+m_{b-1})}\nonumber\\
%&\times&\frac{\theta_{\alpha_a\alpha_b}(\widetilde{t}_{aa+1}+\widetilde{t}_{a+1a+2}+...+\widetilde{t}_{b-1b}+m_a+m_{a+1}+...+m_{b-1}+m_b)}{\theta_{\alpha_a\alpha_b}(\widetilde{t}_{aa+1}+\widetilde{t}_{a+1a+2}+...+\widetilde{t}_{b-1b}+m_{a+1}+...+m_{b-1}+m_b)}\bigg)\nonumber\\
&=& \sum_{\alpha_{1}}Q^{|\alpha_1|}\mathcal{Z}_{(1,M-1)}(\tau, \rho, m_{i}\,,\widetilde{t}_{ab},\epsilon)
\eea
where   $t_{ab}$ and $m_{i}$ do not include the moduli which are tuned to zero.
%\subsection*{Remark :}
More generally and at the same point $Q_i=Q$  in the moduli space we expect similar  structure  for $\mathcal{Z}_{(N,M)}$ 

\bea\label{eq:degMN}
\mathcal{Z}_{(N,M)}(\tau, \rho, m_{a},\widetilde{t}_{ab},\epsilon)&=& \sum_{\alpha_{a}^{i}}\,\prod_{i=1}^{N}Q^{|\alpha^{(i)}|}\,\prod_{i=1}^{N}\prod_{a=2}^{M}\,\frac{\vartheta_{\alpha_{a}^{i+1}\alpha_{a}^{i}}(m_a)}{\vartheta_{\alpha_{a}^{i}\alpha_{a}^{i}}(0)}\nonumber\\
&\times&
\prod_{a=2}^{M-1}\prod_{b=a+1}^M\frac{\vartheta_{\alpha_a^i\alpha_b^{i+1}}(\widetilde{t}_{aa+1}+\widetilde{t}_{a+1a+2}+...+\widetilde{t}_{b-1b}+m_{a+1}+...+m_{b-1})}{\vartheta_{\alpha_a^i\alpha_b^{i}}(\widetilde{t}_{aa+1}+\widetilde{t}_{a+1a+2}+...+\widetilde{t}_{b-1b}+m_a+m_{a+1}+...+m_{b-1})}\nonumber\\
&\times&\frac{\vartheta_{\alpha_a^i\alpha_b^{i+1}}(\widetilde{t}_{aa+1}+\widetilde{t}_{a+1a+2}+...+\widetilde{t}_{b-1b}+m_a+m_{a+1}+...+m_{b-1}+m_b)}{\vartheta_{\alpha_a^i\alpha_b^{i}}(\widetilde{t}_{aa+1}+\widetilde{t}_{a+1a+2}+...+\widetilde{t}_{b-1b}+m_{a+1}+...+m_{b-1}+m_b)}\nonumber\\
&\times&\prod_{i=1}^N\frac{\vartheta_{\alpha_{1}^{i+1}\alpha_{1}^{i}}(m_1)}{\vartheta_{\alpha_{1}^{i}\alpha_{1}^{i}}(0)}\bigg( \frac{\vartheta_{\alpha_1^i\alpha_2^{i+1}}(\widetilde{t}_{12})\vartheta_{\alpha_1^i\alpha_2^{i+1}}(\widetilde{t}_{12}+m_1+m_{2})}{\vartheta_{\alpha_1^i\alpha_2^{i}}(\widetilde{t}_{12}+m_1)\vartheta_{\alpha_1^i\alpha_2^{i}}(\widetilde{t}_{12}+m_{2})}\nonumber\\
&\times&\frac{\vartheta_{\alpha_1^i\alpha_3^{i+1}}(\widetilde{t}_{12}+\widetilde{t}_{23}+m_2)\vartheta_{\alpha_1^i\alpha_3^{i+1}}(\widetilde{t}_{12}+\widetilde{t}_{23}+m_1+m_{2}+m_3)}{\vartheta_{\alpha_1^i\alpha_3^{i}}(\widetilde{t}_{12}+\widetilde{t}_{23}+m_1+m_2)\vartheta_{\alpha_1^i\alpha_3^{i}}(\widetilde{t}_{12}+\widetilde{t}_{23}+m_2+m_3)}
\nonumber\\...
&\times&\frac{\vartheta_{\alpha_1^i\alpha_M^{i+1}}(\widetilde{t}_{12}+\widetilde{t}_{23}+...+\widetilde{t}_{M-1M}+m_1+m_{2}+...+m_{M-1}+m_M)}{\vartheta_{\alpha_1^i\alpha_M^{i+1}}(\widetilde{t}_{12}+\widetilde{t}_{23}+...+\widetilde{t}_{M-1M}+m_{2}+...+m_{M-1}+m_M)}\bigg)\nonumber\\
%\nonumber\\&=&\big( \sum_{\alpha_{1}^{i}}\prod_{i=1}^NQ_i^{\alpha_1^{(i)}}\big) \mathcal{Z}_{(N,M-1)}(\tau, \rho, m_{q},\widetilde{t}_{pq},\epsilon)\nonumber\\
\eea
In the limit $m_1\to 0$
\bea
\mathcal{Z}_{(N,M)}(\tau, \rho, m_{a},\widetilde{t}_{ab},\epsilon)&=&\big( \sum_{\alpha_{1}^{i}}\prod_{i=1}^NQ^{\alpha_1^{(i)}}\big) \mathcal{Z}_{(N,M-1)}(\tau, \rho, m_{q},\widetilde{t}_{pq},\epsilon)\nonumber\\
\eea
Similar recursive structure in (N,M) shows up in the limits  $m_i=0$ (for any i=2,...) or  $\tilde{t}_i=0$.
 From mathematical viewpoint such degenerations have been discussed in \cite{li1998symplectic,liu2005transformation}.

%\section{Physical consequences of the degenerations}\label{physicalcon}
%Recall the following degeneration (\ref{eq:Deg1})
%\bea
%\mathcal{Z}_{(N,M)}(\tau, \rho, t_{a,a+1}=m,\epsilon)=\mathcal{Z}_{(1,1)}(M\,\tau,\rho,M\,m,\epsilon)^{N}
%\eea
%This degeneration corresponds to a $U(M)^N$ quiver gauge theory degenerating to a $U(1)^N$  gauge theory. Moreover the gauge coupling constant $\tau$ and the hypermultiplet mass parameter $m$ are scaled to $M\tau$ and $Mm$ under the degeneration. This rescaling  corresponds to  multiple wrapping number of the D-branes along the
%$\tau$ and $m$ directions.  \\
%Similarly the second degeneration of the $\mathcal{Z}_{N,M}$  (\ref{eq:degMN}) that we discussed and is given by
%\bea
%\mathcal{Z}_{(N,M)}(\tau, \rho, m_{a},t_{ab},\epsilon)&=& \sum_{\alpha_{a}^{i}}F^{\alpha_{a}^{i}}(\tau, \rho, m_{c,d},t_{cd}) \mathcal{Z}^{\alpha_{1}^{i},\alpha_{2}^{i},...,\alpha_{M-1}^{i}}_{(N,M-1)}(\tau, \rho, m_{p,q},t_{pq},\epsilon)\nonumber\\
%\eea
%has an interesting physical interpretation. The limit $m_i\to 0$ corresponds to supersymmetry enhancement to $N=4$ and we get a decoupling factor of $\eta(\tau)$. This is true only for $N\in\mathbb{N},M=1$. For $N\in\mathbb{N},M\in\mathbb{N}_{\ge 2}$ the factorisation is only partial.\\

\section{Discussions }\label{sec:Dis}
The  compactified 5-brane web given in fig.\ref{Fig:WebToric2} gives rise to a five dimensional $\mathcal{N}=2$ supersymmetric gauge theory on the common worldvolume. This 5-branes web can be deformed to include also $(1,1)$ 5-branes. In string theory this is interpreted as the splitting of  the D5-branes on the NS5-brane world volume. In other words the string tension is turned on  for the strings that are stretched between D5-branes.  It gives rise to the mass deformation of the bifundamental hypermultiplets in the five dimensional gauge theory. The mass deformation results in the breaking of supersymmetry to $\mathcal{N}=1$ in five dimensions. Because of the toric compactification of the 5-branes web one gets affine $\hat{A}_{N-1}$ quiver gauge theory with an $SU(N)$ gauge group at each node and one  bifundamental matter stretched between adjacent nodes. There are $M$ coupling constants $\tau_i,i=1,...,M$ for each node such that
 \bea
 \sum_{i=1}^M\tau_i=\frac{1}{R_1}
 \eea
where $R_1$ is the radius of the $S^1$ on which M5-brane theory is compactified. In geometrical terms each gauge coupling constant is related to the area of a distinct curve in CY threefold. If there are more than one, though equivalent, choices of these curves, this gives rise to dual  gauge theory formulations of the same system. In other words for the web of $M$ NS5-branes and $N$ D5-branes  the gauge theory on the D5-branes is given by
\bea\label{eq:quiv1}
\mbox{gauge group} &&: U(1)\times SU(N)_1\times SU(N)_2\times ... \times SU(N)_M\nonumber\\
\mbox{hypermultiplet representation} &&: \oplus_{i=1}^M\bigg( (N_a,\bar{N}_{a+1})\oplus (\bar{N}_a,N_{a+1})\bigg)
\eea
where $N_a$ is the $SU(N)$ fundamental representation of the a-th node and $\bar{N}_a$ the complex conjugate one.
The partition function of the quiver gauge theories given in (\ref{eq:quiv1})   can be computed directly by using Nekrasov instanton calculus as described in \cite{Hohenegger:2013ala}. In doing so one has to take  into account the non-trivial winding of strings on the compact direction transverse to the 5-branes. There is interesting physical interpretations of these degenerations. In the previous sections we have discussed how various degenerations of the mirror curve is related to certain degeneration of the corresponding partition functions $\mathcal{Z}_{(N,M)}$.
%Lately another powerful method of computing the partition function was proposed in  \cite{Haghighat:2013gba} in terms of  M-strings, which are one dimensional intersections of M5 and M2 branes.
%The  table given in figure \ref{figure:coordinates} summarises the coordinate labels  and specifies the world volume directions of BPS M5-M2-M-string configuration.\\
Recall the following degeneration (\ref{eq:Deg1})
\bea
\mathcal{Z}_{(N,M)}(\tau, \rho, t_{a,a+1}=m,\epsilon)=\mathcal{Z}_{(1,1)}(M\,\tau,\rho,M\,m,\epsilon)^{N}
\eea
This degeneration corresponds to a $U(N)^M$ quiver gauge theory degenerating to a $U(1)^M$  gauge theory. Moreover the gauge coupling constant $\tau$ and the hypermultiplet mass parameter $m$ are scaled to $M\tau$ and $Mm$ under the degeneration. This rescaling  corresponds to  multiple wrapping number of the D-branes along the
$\tau$ and $m$ directions.  \\
Similarly the second degeneration of the $\mathcal{Z}_{N,M}$  (\ref{eq:degMN}) that we discussed and is given by
\bea
\mathcal{Z}_{(N,M)}(\tau, \rho, m_{a},t_{ab},\epsilon)&=&\big( \sum_{\alpha_{1}^{i}}\prod_{i=1}^NQ_i^{\alpha_1^{(i)}}\big) \mathcal{Z}_{(N,M-1)}(\tau, \rho, m_{q},\widetilde{t}_{pq},\epsilon)\nonumber\\
\eea
has an interesting physical interpretation. The limit $m_i\to 0$ corresponds to supersymmetry enhancement to $N=4$ and we get a decoupling factor of $\eta(\tau)$. \\
%This is true only for $N\in\mathbb{N},M=1$. For $N\in\mathbb{N},M\in\mathbb{N}_{\ge 2}$ the factorisation is only partial.\\

\section{Conclusions}\label{sec:Con}

%This is true only for $N\in\mathbb{N},M=1$. For $N\in\mathbb{N},M\in\mathbb{N}_{\ge 2}$ the factorisation is only partial.\\
This paper explored some interesting consequences  of the mirror symmetry of the local CY threefold $X_{N,M}$.
We investigated some interesting properties of the type $A$ topological string partition function of $X_{N,M}$  in   special regions of the K\"ahler moduli space. We have called these degenerate limits, because in these limits the partition functions  on $X_{N,M}$ collapse to those on $X_{N,M-1}$ in various ways.  In accordance with mirror symmetry the degeneration behaviour on the type A side is reproduced on the type B side in the degeneration of the   mirror curves into lower genus curves.\\
% We showed that appearance of quasi Siegel modular forms
%in the description of Gromov-Witten invariants. The Nekrasov-Shatashvili limit played an important role in the computation as has been discussed thoroughly in \cite{Hohenegger:2016eqy}.\\
For future directions it would be interesting to study the analogous properties of $\mathcal{Z}_{N,M}$ and quantum mirror curves for the general $\Omega$-background .i.e. $\epsilon_1\ne 0$ and/or $\epsilon_1\ne 0$ and $\epsilon_1\ne\epsilon_2$ and at an arbitrary point of the K\"ahler moduli space of $X_{N,M}$. It will also be interesting to study the modular properties of the free energy $\log(\hat{\mathcal{Z}}_{(N,M)}(\tau, \rho, \epsilon_{}, m, \mbox{\bf{t}}))$ and the single particle free energy \cite{Hohenegger:2015cba} $PLog(\hat{\mathcal{Z}}_{(N,M)}(\tau, \rho, \epsilon_{}, m, \mbox{\bf{t}}))$  along the lines of \cite{Hohenegger_2020}. It is also interesting to generalise the quantisation of classical DELL system as done in \cite{Koroteev:2019gqi} to the case where the underlying abelian variety has (M,N) polarization.
%Also see recent work \cite{Jafarzade:2017fsc,Chen:2020jla}.
\section*{Acknowledgement}
The authors would like to thank  Amer Iqbal for  discussions and acknowledge the support of  the Abdus Salam School of Mathematical Sciences, Lahore,Pakistan.
\begin{appendix}
\section{\hspace{0.9cm}\\Geometry of $X_{N,M}$: a quick review}\label{section:XMNGeo}
 The non-compact CY 3-fold $X_{1,1}$ is defined as the partial compactification \cite{Kanazawa:2016tnt,Hohenegger:2013ala} of the resolved conifold geometry. The later is given by  $\mathbb{C}^{\times}\times \mathbb{C}^{\times}$ fibered over the $z$-plane. The partial compactification is achieved  by compactifying each of the two $\mathbb{C}^{\times}$ fibers to a $\mathbb{T}^2$ fiber.  Of the three K\"ahler parameters $\tau,\rho,m$ of the CY 3-fold $X_{1,1}$, $\rho$ and $\tau$ correspond to the elliptic fibers and $m$ corresponds to the curve class of the exceptional $\mathbb{P}^1$ of the resolved conifold.
We will define the non-compact CY 3-fold $X_{N,M}$ for $N,M\in\mathbb{N}$ as the $\mathbb{Z}_N\times\mathbb{Z}_M$ orbifold of $X_{1,1}$.\\
In toric geometry the equation of the conifold  given by
\bea
z_1z_2-z_3z_4=0,\quad z_1,z_2,z_2,z_4\in\mathbb{C}
\eea
is translated to an  equation on integer latices parametrised by 3-vectors $v_1,v_2,v_3,v_4$
\bea
v_1+v_2-v_3-v_4=0.
\eea
The CY condition constrains the geometry to a plane.
 The irreducible toric rational curves of  the 2-dimensional cone are given by
\bea
C^1_{(a,b)}:&=&\mathbb{R}_{\ge 0}Conv(\{(a+1,b,1),(a,b+1,1)\}),\quad  C^2_{(a,b)}:=\mathbb{R}_{\ge 0}Conv(\{(a,b,1),(a,b+1,1)\}),\nonumber\\
C^3_{(a,b)}:&=&\mathbb{R}_{\ge 0}Conv(\{(a,b,1),(a+1,b,1)\}).
\eea
for all $a,b\in\mathbb{Z}$. The K\"ahler variables $q_i$ corresponding to $C^i$ are defined as the exponential of the symplectic area of $C^i$. The author in \cite{Kanazawa:2016tnt} computes Strominger-Yau-Zaslow (SYZ) \cite{Strominger:1996it} mirror of  the local CY3-fold $X_{N,M}$, which is given by
\bea
uv=\sum_{a,b=0}^{N-1,M-1}\Delta_{a,b}\sum_{c,d\in\mathbb{Z}^2}q^{C_{(cN+a,dM+b)}}z_1^{cN+a}z_2^{dM+b}
\eea
where $\Delta_{a,b}$ encodes the data of  the open Gromov-Witten invariants, $z_1,z_2$ are coordinates of the abelian variety of polarisation $(N,M)$, and $u,v$ are the sections of certain line bundles on the abelian variety. The zero locus
\bea
\sum_{a,b=0}^{N-1,M-1}\Delta_{a,b}\sum_{c,d\in\mathbb{Z}^2}q^{C_{(cN+a,dM+b)}}z_1^{cN+a}z_2^{dM+b}=0
\eea
defines a curve with genus $NM+1$ with $(N,M)$ polarisation. For illustration, consider the CY3-fold $X_{1,1}$, for which the cone of effective curves is given by $\mathbb{R}_{\ge0}\{C^1,C^2,C^3\}$. To make the modularity of the system manifest, we redefine the curve classes as
\bea
C_{\tau}=C^1+C^2,\quad C_{\rho}=C^1+C^3,\quad C_{\sigma}=C^1
\eea
for which the corresponding K\"ahler parameters are denoted as $q_{\tau}=q_1q_2=e^{2\pi i\tau},q_{\rho}=q_1q_3=e^{2\pi i\rho},q_{\sigma}=q_1=e^{2\pi i\sigma}$. Then following the SYZ program, the SYZ mirror of $X_{1,1}$ is given by
\bea
uv=\Delta(q)\sum_{c,d\in\mathbb{Z}^2}q^{C_{(c,d)}}z_1^{c}z_2^{d}
\eea
Moreover it turns out that the right hand side can be re-written in terms of theta function as
\bea
uv=\Delta(\Omega)\Theta_2  \left[ {\begin{array}{cc}
  0 \\
  (-\frac{\tau}{2},-\frac{\rho}{2}) \\
  \end{array} } \right](z_1,z_2;\Omega)
\eea
where $\Theta_2$ is the genus 2 theta function and $\Omega=\begin{pmatrix}N\tau & {\bf \sigma}\\ {\bf \sigma}& M\rho\end{pmatrix}$ is the period matrix of the following genus 2 curve
\bea
\Theta_2  \left[ {\begin{array}{cc}
  0 \\
  (-\frac{\tau}{2},-\frac{\rho}{2}) \\
  \end{array} } \right](z_1,z_2;\Omega)=0
\eea

Moreover the curve classes $C^i$ satisfy the following relations 
\bea
C^1_{(a-1,b)}+C^3_{(a-1,b)}=C^1_{(a,b-1)}+C^3_{(a-1,b)},\nonumber\\
C^1_{(a-1,b)}+C^2_{(a,b)}=C^1_{(a,b-1)}+C^2_{(a,b-1)}.
\eea
%The cone of effective curve classes for  $X_{1,1}$ is given by $\mathbb{R}_{\ge 0}\{C_1,C_2,C_3\}$. The indices $a,b$ are implicit. Due to the  modular properties of $\mathcal{Z}_{1,1}$ the classes $C_1,C_2,C_3$ appear in combinations $C_1+C_2:=C_{\tau},C_1+C_3:=C_{\rho},C_{m}:=C_1$ with corresponding K\"ahler parameters $\tau,\rho,m$ respectively. \\
 For  the local CY 3-fold $X_{N,M}$ a modular covariant basis of generators can be given by
 \bea\label{eq:relations}
 C_{m,(a,b)}&=&C^1_{(a,b)},\quad  C_{\tau,(a,b)}=C^1_{(a,b)}+C^2_{(a,b)},\nonumber\\
 C_{\rho,(a,b)}&=&C^1_{(a,b)}+C^3_{(a,b)}
 \eea
 where $a,b\in\mathbb{Z}$.
 In the \mbox{\it{fundamental  domain}} of the $(N,M)$-web there are $3MN$ toric rational curves where $a\in\mathbb{Z}_N,b\in\mathbb{Z}_M$. Due to the $2NM$ constraints in  (\ref{eq:relations}) and torus periodicity the effective rank is $MN+2$.
 
\section{\hspace{0.9cm}\\$\sum_{a=0}^{N-1}\bf{m}_{a,b}$ \mbox{is independent of }$b$: proof}\label{sec:mabproof}
Here we prove the identity used in subsection \ref{secXMN}.\\
Note that in our notation the curve classes $C^1_{(a,b)}$ are represented by the K\"ahler parameters  $\bf{m}_{a,b}$. 
Using the first relation in eq.(\ref{eq:relations}), we can write the following summation 
\bea
\sum_{a=0}^{p-1}(C^1_{(a-1,b)}+C^3_{(a-1,b)})=\sum_{a=0}^{p-1}(C^1_{(a,b-1)}+C^3_{(a-1,b)}),
\eea
Due to the compactification of web diagram on a torus there is periodicity relation $C_{(-1,b)}^1=C^1_{(p-1,b)}$.
After simplification the second term cancels on both sides  and we get 
\bea
\sum_{a=0}^{p-1}(C^1_{(a-1,b)})=\sum_{a=0}^{p-1}(C^1_{(a,b-1)}),
\eea
Expanding  the left side
\bea
\sum_{a=0}^{p-1}(C^1_{(-1,b)}+C^1_{(0,b)}+C^1_{(1,b)}+...+C^1_{(p-3,b)}+C^1_{(p-2,b)})=\sum_{a=0}^{p-1}(C^1_{(a,b-1)}),
\eea
 Rearranging the terms after using Using $C_{(-1,b)}^1=C^1_{(p-1,b)}$, we  obtain the desired relation 
% \cite{Bousseau:2020ckw,kawai2000string,lockhart2012superconformal,Huang:2011qx,Hellerman_2012,Gopakumar:1998jq,Gopakumar:1998ii,huang2010direct,Aharony_2000,nekrasov2002seibergwitten,Iqbal_2009,Aganagic_2004,Bershadsky_1996,leung1997branes,Aharony_1998,li1998symplectic,liu2005transformation,katz1997mirror,Bhardwaj_2016,Haghighat_2014,Marshakov_2000}
\bea
\sum_{a=0}^{p-1}C^1_{(a,b)}=\sum_{a=0}^{p-1}C^1_{(a,b-1)} \quad .
\eea
\end{appendix}

\bibliographystyle{JHEP}
%\nocite{*}
\bibliography{bibliography}
\end{document}